\documentclass[reprint,aps,prb,amsmath,amssymb,twocolumn,showkeys, superscriptaddress,floatfix]{revtex4-2}
\pdfoutput=1
\usepackage{graphicx}
\usepackage{dcolumn}
\usepackage{bm}
\usepackage[colorlinks = true, allcolors = blue]{hyperref}
\usepackage{verbatim}
\usepackage{soul}
\usepackage{float}
\usepackage{xcolor}
\usepackage{multirow}
\usepackage{ulem}
\DeclareMathOperator{\tr}{tr}

\begin{document}
\title{Twisted bilayered graphenes at magic angles and Casimir interactions: correlation-driven effects}
\author{Pablo Rodriguez-Lopez}
\email{pablo.ropez@urjc.es}
\affiliation{{\'A}rea de Electromagnetismo and Grupo Interdisciplinar de Sistemas Complejos (GISC), Universidad Rey Juan Carlos, 28933, M{\'o}stoles, Madrid, Spain}

\author{Dai-Nam Le}
\email{dainamle@usf.edu}
\affiliation{Department of Physics, University of South Florida, Tampa, Florida 33620, USA}
\affiliation{Atomic~Molecular~and~Optical~Physics~Research~Group, Advanced Institute of Materials Science, Ton Duc Thang University, Ho~Chi~Minh~City 700000, Vietnam}

\author{Mar\'{i}a J. Calder\'{o}n}
\email{calderon@icmm.csic.es}
\affiliation{Instituto de Ciencia de Materiales de Madrid (ICMM), Consejo Superior de Investigaciones Científicas (CSIC), Sor Juana Inés de la Cruz 3, 28049 Madrid, Spain.}

\author{Elena Bascones}
\email{leni.bascones@csic.es}
\affiliation{Instituto de Ciencia de Materiales de Madrid (ICMM), Consejo Superior de Investigaciones Científicas (CSIC), Sor Juana Inés de la Cruz 3, 28049 Madrid, Spain.}

\author{Lilia M. Woods} 
\email{lmwoods@usf.edu}
\thanks{Corresponding author.}
\affiliation{Department of Physics, University of South Florida, Tampa, Florida 33620, USA}

\date{\today}

\begin{abstract}
Twisted bilayered graphenes at magic angles are systems housing long ranged periodicity of moir{\'e} patterns together with short ranged periodicity associated with the individual graphenes. Such materials are a fertile ground for novel states largely driven by electronic correlations. Here we find that the ubiquitous Casimir force can serve as a platform for macroscopic manifestations of the quantum effects stemming from the magic angle bilayered graphenes properties and their phases determined by electronic correlations. By utilizing comprehensive calculations for the electronic and optical response, we find that Casimir torque can probe anisotropy from the Drude conductivities in nematic states, while repulsion in the Casimir force can help identify topologically nontrivial phases in magic angle twisted bilayered graphenes. 

\end{abstract}

\maketitle

\section{Introduction}

A twisted bilayered graphene (TBG) is a moir{\'e} superlattice material consisting of two graphene sheets whose stacking is quantified by a twist angle $\theta$ of the relative orientation of the crystal axis for each graphene \cite{Andrei2020}. In the case of special {\it magic} $\theta$, the occurrence of very long superlattice period of AA-AB stacked domains coupled with the much shorter ranged atomistic periodicity of the individual graphenes creates an environment for unprecedented properties at the nanoscale \cite{Nimbalkar2020}.  A number of striking experimental observations have shown that such TBGs can have insulating, superconducting, and topological states as doping is slightly changed \cite{Lu2019,Sharpe2019,Jiang2019,Liu2021,Cao2018,Cao2018-1}. These phases are directly connected with the unique band structure with flat bands around the charge neutrality point (CNP), which also indicate the much enhanced role of electron-electron correlations \cite{Morell2010,Bistritzer2011,dosSantos2012}. 

Our current understanding is that electronic interactions lead to the breaking of various symmetries, considered to be an inherent reason for the emergence of the various states of magic angle TBGs \cite{Xie2020,Po2018}. Among these phases, STM measurements show evidence of nematicity due to broken $C_3$ symmetry as the chemical potential lies within the flat bands of the TBG \cite{Choi2019,Jiang2019,Xie2019,Calderon2020,Cao2021}. Signatures of gap opening
at the CNP have been detected in transport properties, which may be due to broken $C_2T$ symmetry \cite{Lu2019,Stepanov2020}.  While much research is currently directed towards the electronic properties, the dielectric response of TBG at magic angles has received limited attention. The optical conductivity is a direct manifestation of the underlying band structure and recent theoretical studies have shown that nematicity at charge neutrality can produce a metal Lifshitz transition which results in {\it anisotropic} Drude conductivities \cite{Calderon2020}. Such anisotropy is present in iron based superconductors as well, as revealed by extensive experiments \cite{Blomberg2013,Chu2010,Ishida2013,Valenzuela2010}.

The optical response for the different phases of TBGs at magic angles near half filling will have profound effects on their light-matter interactions. One example is the universal Casimir force, which exists between any two types of objects \cite{Woods2016,Klimchitskaya2009}. Despite its ubiquitous nature, a diverse dependence upon distance, sign, or characteristic constants is found when the material-dependent optical response is taken into account. Understanding the Casimir force is important for the basic physics of the quantum vacuum. It also has significant practical implications especially in the context of the stability of weakly bound materials or for the operation of micro-machines, where stiction, adhesion, and friction phenomena become relevant \cite{Yapu2003,Kardar1999}.   

The Casimir force arises from electromagnetic fluctuations exchanged between objects and it is a remarkable macroscopic manifestation of quantum physics. Recent theoretical models and experiments have shown that the Dirac-like states in graphene lead to dominant thermal fluctuations at much smaller separations when compared to the $\mu$m range for typical metals and dielectrics \cite{Liu2021B,Mohideen2021,Khusnutdinov2018}. It has also been suggested that repulsive and quantized Casimir interactions may be possible in 2D Chern insulators due to their anomalous Hall conductivity associated with an integer Chern number, but in 3D Weyl semimetals the nontrivial topology plays a secondary role \cite{Rodriguez-Lopez2014,Rodriguez-Lopez2020,Fialkovsky2018,Babamahdi2021}. Another important observation is that for optically anisotropic materials, the exchanged electromagnetic fluctuations experience different refractive indices along the different directions causing the emergence of a Casimir torque \cite{Broer21,Antezza2020,Lu2016}. Recent experiments have demonstrated the relative rotation of bifringent and liquid crystals showing that the separation distance and choice of materials may be effective control knobs \cite{Somers2018}.

It is clear that fluctuation induced interactions depend strongly on the fundamental properties of materials. In fact, Casimir-related phenomena can be used as an effective means to observe signatures of fundamental properties originating from their underlying atomic and electronic structure. Dirac-like physics, nontrivial topology, and anisotropy have specific fingerprints in Casimir phenomena \cite{Rodriguez-Lopez2017,Farias2020,Lu2021}. This basic knowledge is instrumental for the design of better nano and micro-devices, which are relevant for several technologies, as discussed earlier \cite{Palasantzas2020,Javor2021}.

In this study, we investigate the electronic and optical response properties of TBGs at magic angles showing a unique range of Casimir phenomena. Possible breaking of $C_3$, $T$, or $C_2$ symmetries due to electron correlations give rise to topologically nontrivial and nematic phases near half filling. We show that small changes in the electronic structure upon such symmetry breaking conditions lead to significant changes in the optical response, which in turn have profound effects on the Casimir interaction. It is remarkable that the nematic state of magic angle TBGs due to broken $C_3$ symmetry results in anisotropic Drude conductivities giving rise to Casimir torque, while the topologically nontrivial TBG due to broken $T$ symmetry can experience Casimir repulsion. While most studies are pursuing methods for direct electronic or transport evidence of magic angle TBGs, our comprehensive investigation shows that Casimir interaction phenomena are another set of means to probe and distinguish between their different states. 

\section{TBG at magic angles and its properties}
\begin{figure}[ht]
    \centering
    \includegraphics[width = 0.9 \columnwidth]{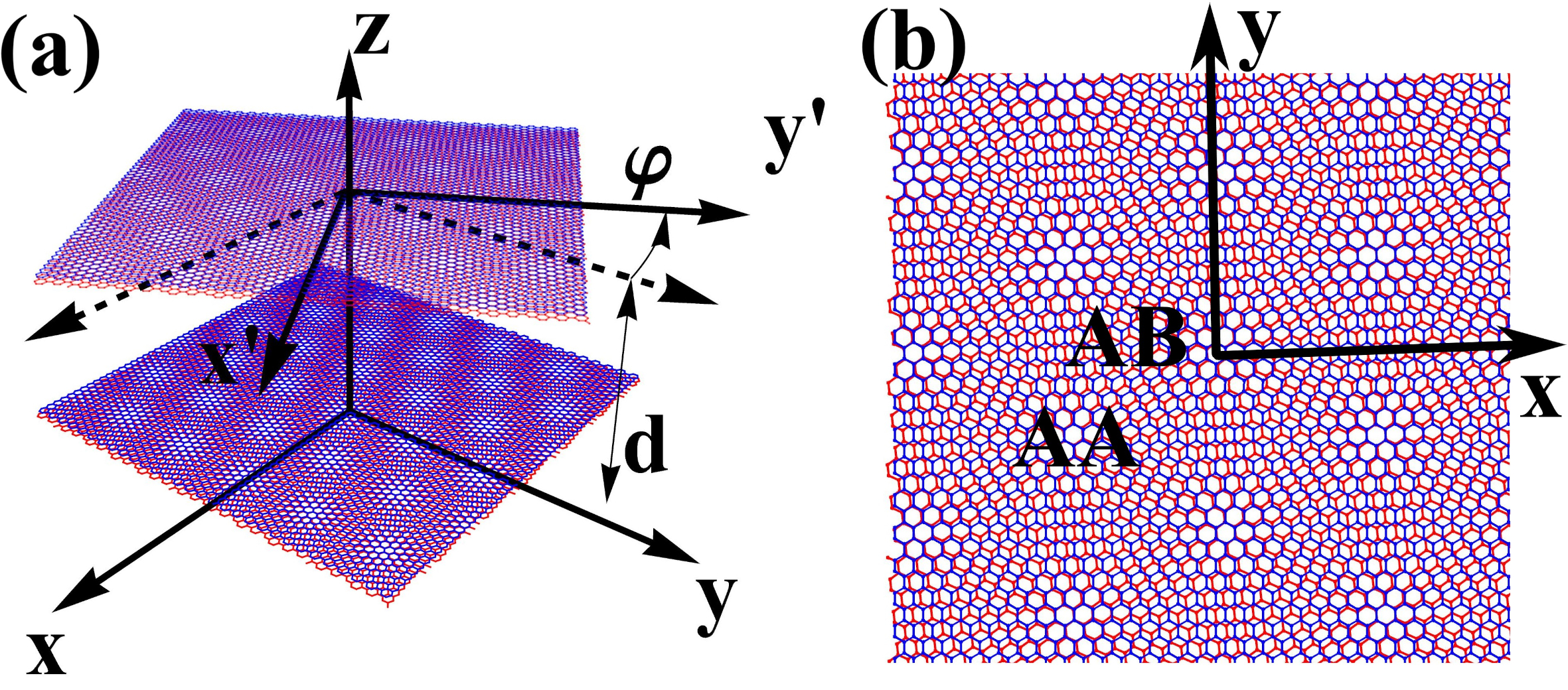}
    \caption{(a) Two identical parallel TBGs in the $x-y$-plane separated by a distance $d$ along the $z$-axis. The relative orientation of their optical axes is denoted by the angle $\varphi$; (b) a single TBG whose optical axes are along the $x, y$-directions with AA and AB stacking regions.}
    \label{fig:1}
\end{figure}

We consider the Casimir interaction between a pair of identical TBGs separated by a distance $d$ along the $z$-axis, as shown in Fig. \ref{fig:1}a. The relative orientation of the optical axes of each TBG is denoted by $\varphi$.  Each TBG consists of two graphene layers with a relative angle $\theta$ of their optical axes and Fig. 
\ref{fig:1}b shows a top view of a single TBG with its AA-AB stacking pattern. In what follows we focus on a fully relaxed (FR) TBG with $\theta_{FR}=0.9^{\circ}$. 

The single particle TBG electronic structure is taken within an effective moir\'e ten-orbital tight binding model for each valley and spin degree of freedom \cite{Carr2019,Po2019}. Each of the valleys, assumed to be uncoupled, correspond to one of the two Dirac cones of graphene. The band structure results from the fitting to self-consistent {\it ab initio} 
$k\cdot p$ calculations which account for both, out of plane relaxation of the AA and AB domains and the in-plain strain of individual graphenes due to the stacking \cite{Carr2019}. This model yields  relatively flat bands near the CNP and it  satisfies the underlying symmetry of the system including the invariance under mirror flip with respect to the $y$-axis ($M_{2y}$), rotation with respect to axis perpendicular to the TBG ($C_3$), and the product ($C_2T$) of rotation with an axis perpendicular to the TBG $(C_2)$ and time-reversal symmetry ($T$) within each  valley. Details of the model are given in Methods of Calculations and the Supplemental Information. 

This effective ten-orbital model allows introducing phenomenologically at the mean field level different symmetry breaking orders which mimic signatures observed experimentally. Below we focus on the breaking of $C_2T$ symmetry, introduced through a parameter $\alpha_{\nu}$, with $\nu$ labelling the valley. We also consider nematic orders breaking the $C_3$ symmetry, which are introduced in two different ways through $\eta$ and $\beta$ parameters. The order parameters $\alpha_\nu, \beta$, and $\eta$ are obtained self-consistently from the microscopic model in ~\cite{Choi2019,Calderon2020}. For more details, see Methods of Calculations and the Supplementary Information. The values of $\alpha_{\nu}$, $\beta$ and $\eta$ are given in meV.

\begin{center}
\begin{figure}[ht]
    \centering
    \includegraphics[width = 0.95 \textwidth]{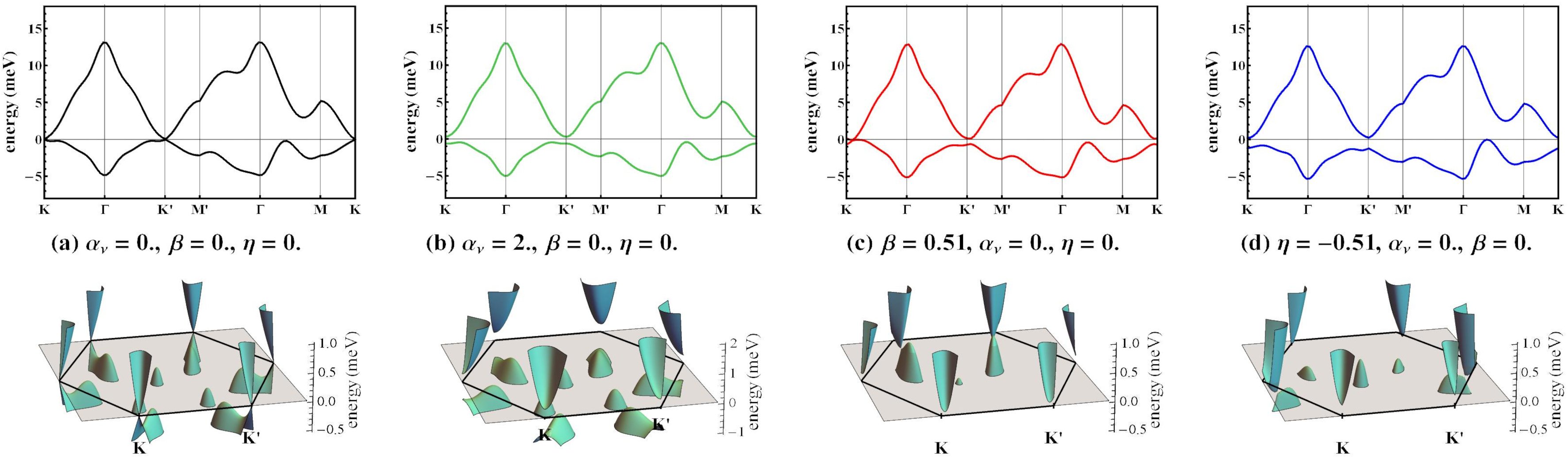}
    \caption{The low energy band structure for a single TBG with a $\theta_{FR}=0.9^{\circ}$ twist angle is given as calculated with the ten-band fully relaxed model \cite{Calderon2020,Carr2019}. The CNP corresponds to $n=5.000$. The top panels show the bands along high symmetry directions and the bottom ones give a zoom-in around the Fermi level in the whole Brillouin zone for a: (a) non-correlated state with $\alpha_\nu=\beta=\eta=0$; (b) $C_2T$ symmetry breaking state with $\alpha_\nu=2$; (c) nematic state with $\beta=0.51$ breaking the $C_3$ symmetry; (d)  nematic state with $\eta=-0.51$ breaking the $C_3$ symmetry.}
    \label{fig:2}
\end{figure}
\end{center}

In Fig. \ref{fig:2}, we show representative cases with specific numerical values of the $\alpha_\nu, \beta, \eta$ parameters for the distinct phases near half filling of a single TBG at $\theta_{FR}=0.9^{\circ}$. For $\alpha_\nu=\beta=\eta=0$ all inherent symmetries are preserved. The CNP corresponds to half-filling with $n=5.00$ occupation number, such that the Fermi level passes through the Dirac $K, K'$ points of the moir\'e mini-Brillouin zone signalling a semimetallic behavior. This is also 
seen in the three-dimensional band structure in Fig. \ref{fig:2}a. For $\alpha_\nu \neq 0$, the two valleys become uncoupled and the broken $C_2T$ symmetry is responsible for shifting the bands in energy and opening a gap at the Fermi level as shown in Fig. \ref{fig:2}b. Such a band structure can be obtained by either  taking  $\alpha_{\nu}=-\alpha_{\bar \nu}=2$, which breaks the $C_2$ symmetry, or by taking $\alpha_{\nu}=\alpha_{\bar \nu}=2$ for which the $T$ symmetry is broken.

Breaking the $C_3$ symmetry, on the other hand, unpins the Dirac points by moving them away from the $K, K'$ points \cite{Po2018}. The band structures corresponding to $\beta=0.51$ and $\eta=-0.51$ are shown in Figs.~\ref{fig:2}(c) and(d). For the nematic cases with the considered $n=4.995$ and $5.005$, small pockets cross the Fermi energy at the CNP [16] and slightly away from it, as shown in Fig. \ref{fig:2}c and d. \cite{Calderon2020}. In Fig. S-1 in the Supplementary Information, the band structure is given in a larger energy window, which depicts that the changes due to various symmetry breaking conditions happen in a small energy range. Nevertheless, the electronic phases with such symmetry breakings are relevant experimentally, as previously discussed \cite{Jiang2019,Lu2019,Xie2019,Stepanov2020}.

\onecolumngrid
\twocolumngrid

\begin{figure}[ht]
    \centering
    \includegraphics[width = 0.95 \columnwidth]{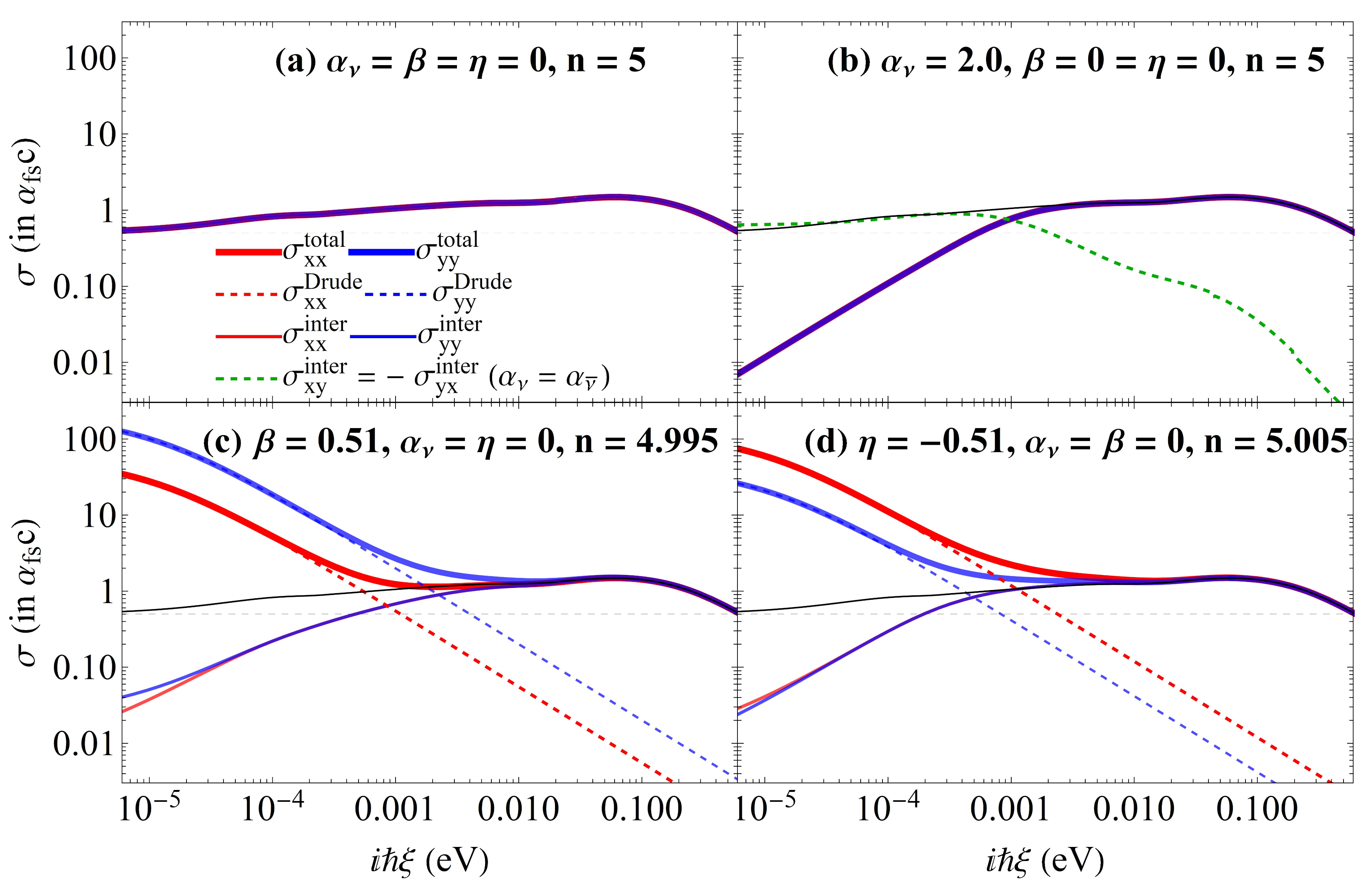}
    \caption{Optical conductivity components of a TBG with a twist angle $\theta_{FR}=0.9^{\circ}$ for a: (a) non-correlated $\alpha_\nu=\beta=\eta=0$ state with all inherent symmetries intact; (b) insulating state with $\alpha_\nu=\alpha_{\bar \nu}=2$; (c) nematic state with $\beta=0.51$; (d) nematic state with $\eta=-0.51$. The thin black curve in (b,c,d) shows the diagonal optical conductivity for the non-correlated state in (a) as a reference.}
    \label{fig:3}
\end{figure}

The optical response of magic angle TBG reflects its unique electronic structure and by using the Kubo formalism within linear response \cite{Kubo1957,Kubo1957-1,Valenzuela2013} the optical conductivity is also calculated numerically (details in Methods of Calculations). Since Casimir interactions are typically considered in an imaginary frequency domain \cite{Abrikosov}, in Fig. \ref{fig:3} the components of the 2D conductivity tensor are shown as a function of $(i\hbar\xi)$ for the TBG states from Fig. \ref{fig:2}.  The optical response for the non-correlated $\alpha_\nu=\beta=\eta=0$ state is found to be isotropic with nearly constant interband components $\sigma_{xx}=\sigma_{yy}\approx 2\sigma_g$ with $\sigma_g=e^2/4\hbar$ being the universal conductivity of a single graphene (Fig. \ref{fig:3}a). Thus, when all symmetries are preserved, the TBG is optically equivalent to two individual graphene monolayers \cite{Stauber2013} due to the nearly frequency independent interband conductivity components at very low frequencies.

In the case of a gapped TBG ($\alpha_\nu \neq 0$), in addition to the interband diagonal conductivity, there is a topological Hall conductivity associated with each valley for which the Chern number takes the sign for its $\alpha_{\nu}, \alpha_{\bar \nu}$ parameters. For $\alpha_{\nu}=\alpha_{\bar \nu}$, the contributions from both valleys are of equal magnitude and same sign adding up to $\sigma_{xy}=\frac{c\alpha_{\rm fs}}{2\pi}\mathcal C$ at low frequency  (Fig. \ref{fig:3}b). Here, $\alpha_{\rm fs}=\frac{e^{2}}{\hbar c}\approx \frac{1}{137}$ is the fine structure constant and $\mathcal C=2$ is the Chern number for the TBG \cite{Rodriguez-Lopez2017,Ezawa2012}. However, when 
$\alpha_{\nu}=-\alpha_{\bar \nu}$, the contributions from both valley exactly cancel out. Thus we find that a magic angle TBG with a broken $C_2$ symmetry has only interband diagonal conductivity, while the TBG with a broken $T$ symmetry displays a Hall response as well. In both cases, the response for the gapped TBG is isotropic with $\sigma_{xx}=\sigma_{yy}$. The origin of $\sigma_{xy}$ is directly related to the anomalous quantum Hall effect (QAHE) that TBGs  can support \cite{Xie2020,Po2018}. The same topologically nontrivial phase has been found in graphene and graphene-like materials with staggered atomic structure \cite{Ezawa2012,Ezawa2013}, such as silicene and germanene.  The QAHE is induced by external laser and static electric fields in these graphene-like monolayers, while in TBGs studied here, this is an intrinsic effect driven by correlations.  

The response of nematic TBGs is quite different.  Since the $C_3$ rotational symmetry is broken with $\beta \neq 0$ or $\eta \neq 0$, the Fermi pockets in the band structure (see Fig.~\ref{fig:2}c,d) are responsible for the emergence of low frequency optical anisotropy primarily due to the intraband Drude contributions \cite{Calderon2020}. Fig. \ref{fig:3}c,d shows the different Drude terms along the $x$ and $y$ axis for the $\beta=0.51$ and $\eta=-0.51$ nematic parameters, which directly reflects the prominent intraband transitions due to bands crossing the Fermi level. This behavior is similar to the situation in iron superconductors, where nematicity from strong correlations results in dissimilar Drude terms along otherwise equivalent directions \cite{Fernandes2011, Valenzuela2010}. A much smaller anisotropy is found in the interband contributions along the $x$ and $y$ directions in the low frequency range. At larger $(i\hbar\xi)$ the interband transitions dominate the response, and $\sigma_{xx}$ and $\sigma_{yy}$ are practically the same. 

\begin{figure}[ht]
    \centering
   \includegraphics[width = 0.95 \columnwidth]{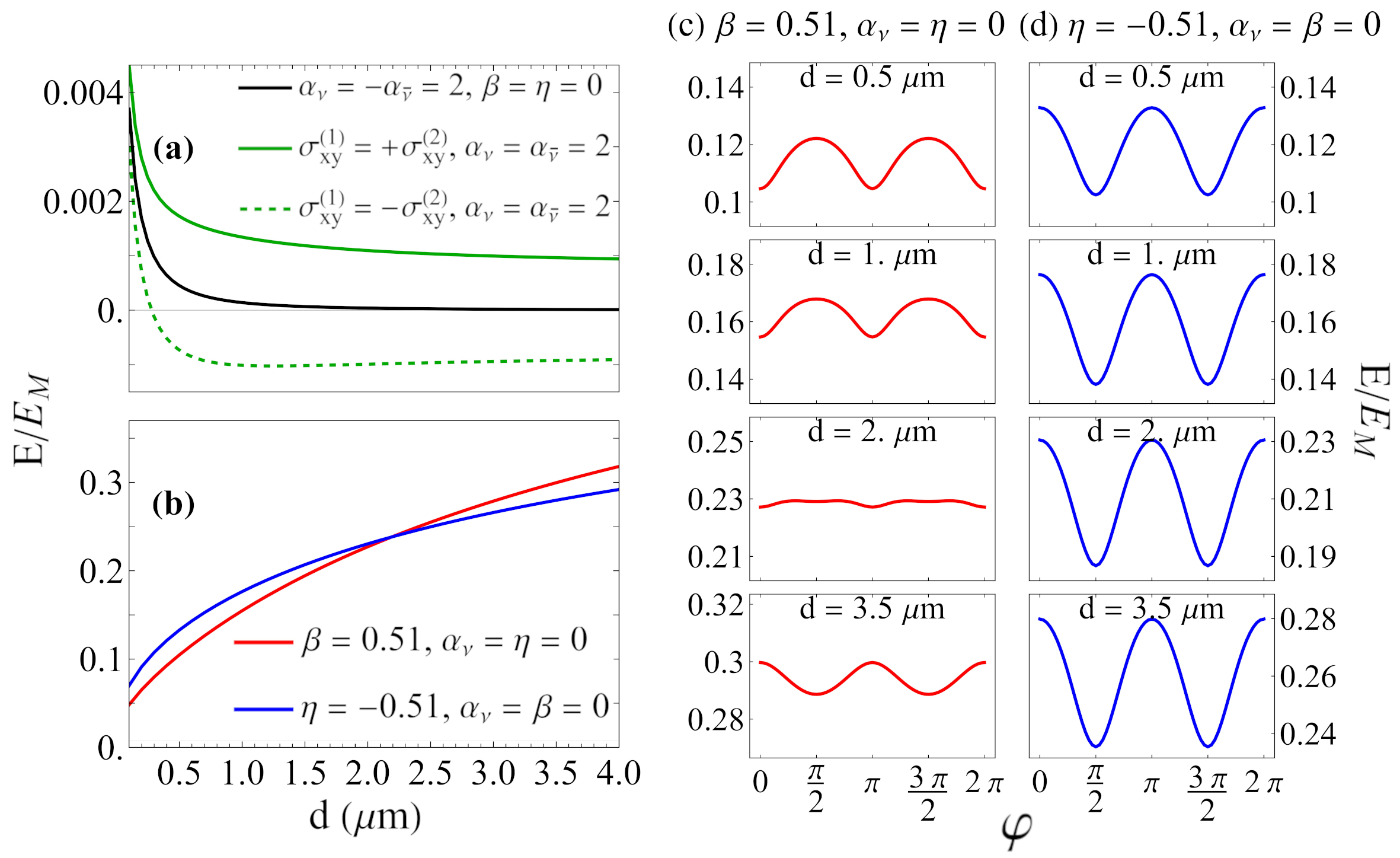}
    \caption{Casimir interaction energy $E$ between two identical TBGs normalized to the Casimir energy between two perfect metals $E_M=-\frac{\pi^2\hbar c}{720 d^3}$ for: (a) $\alpha_\nu=\alpha_{\bar \nu}=2$ when the Hall conductivities and corresponding Chern numbers have the same signs (full green) and when the Hall conductivities and corresponding Chern numbers have opposite signs (dashed green). The black curve corresponds to $\alpha_\nu=-\alpha_{\bar \nu}=2$ when the Chern number is zero; (b) nematic states with $\beta=0.51$ and $\eta=-0.51$ parameters with aligned optical axes, $\varphi=0$. The $E/E_M$ ratio as a function of the angle $\varphi$ between the TBG optical axes at several distances for: (c) $\beta=0.51$ and (d) $\eta=-0.51$.}   
    \label{fig:4}
\end{figure}

\section{Casimir Interactions and Torques}

The Casimir interaction between TBGs can now be calculated using the Lifshitz formalism in the quantum mechanical limit as described in Methods of Calculations, where the electromagnetic boundary conditions and optical response properties are taken into account via the Fresnel reflection matrices. The long ranged interaction energy per unit area of identical noncorrelated TBGs is found as 
\begin{equation}
\label{eqn:1}
E=-\frac{\hbar c \alpha_{\rm fs}}{16\pi d^3}=2E_g,
\end{equation}
which is simply twice the Casimir energy between two graphene monolayers $E_g=-\frac{\hbar c \alpha_{\rm fs}}{32\pi d^3}$ \cite{Drosdoff2010,Drosdoff2012}. The above result is not surprising, since the noncorrelated TBG is a semimetal with Dirac points at the $K, K'$ points with a nearly constant optical conductivity at low frequencies (Fig. \ref{fig:3}) doubled the one of monolayer graphene \cite{Tabert2013,Stauber2013,Calderon2020}. 

The consequences of the parameter $\alpha_{\nu}$ in the Casimir energy are shown in Fig. \ref{fig:4}a. When  $T$ symmetry is broken with $\alpha_{\nu}=\alpha_{\bar \nu}=2$, the gapped TBG has anomalous Hall conductivity which signals that the TBG is effectively a Chern insulator \cite{Ezawa2012,Ezawa2013,Xie2020,Po2018} with $\sigma_{xy}(i\xi=0)=\frac{c \alpha_{\rm fs}}{2\pi}\mathcal C$ with the Chern number $\mathcal C=2$. While at smaller separations the isotropic interband diagonal conductivity dominates the Casimir energy, at larger $d$ the interaction is determined by the Chern numbers of the two TBGs according to $E= -\frac{\hbar c \alpha_{\rm fs}^2}{8\pi^2 d^3}{\mathcal C}_1{\mathcal C}_2$. Since ${\mathcal C}_1, {\mathcal C}_2$ can either be positive or negative, we find an attractive (${\mathcal C}_1{\mathcal C}_2>0$, full green curve) and repulsive (${\mathcal C}_1{\mathcal C}_2<0$, dashed green curve) asymptotic Casimir interaction in Fig. \ref{fig:4}a. Switching the sign of the Hall conductivity for the system in Fig. \ref{fig:1}a can be achieved by flipping one of the TBG upside down. When $C_2$ symmetry is broken with 
$\alpha_{\nu}=-\alpha_{\bar \nu}=2$, the TBG is essentially a quantum spin Hall insulator with $\mathcal C=0$. In that case, we find that the interaction is attractive with an asymptotic behavior $E\sim -\frac{\alpha_{\rm fs}^2}{d^5}$ \cite{Rodriguez-Lopez2014,Rodriguez-Lopez2017}.

In Fig. \ref{fig:4}b, we also show $E$ as a function of TBG separation  for nematic orders with $\beta \neq 0$ or $\eta \neq 0$ with aligned optical axes ($\varphi=0$). It appears  that the interaction between nematic TBGs is attractive and the energy is on the rise as it approaches the one for perfect metals at sufficiently large separations. This type of behavior is due to the balance between strong Drude optical response and interband contributions, also previously found in other 3D topological materials \cite{Rodriguez-Lopez2020}. We further note that due to the $D_{xx}\neq D_{yy}$ anisotropy, the Casimir energy is expected to depend on the relative orientation of the optical axes of the nematic TBGs. In Fig. \ref{fig:4}c,d we show how such a dependence unfolds for the two nematic states. While for $\eta=-0.51$, $E/E_M$ exhibits the same oscillatory like behavior at all shown separations, for $\beta=0.51$ the energy oscillations strongly depend on $d$ with a changing oscillatory pattern.

The $E(\varphi)$ dependence is a fingerprint of a Casimir torque between objects with anisotropic optical response \cite{Barash1978}. This phenomenon has been investigated theoretically in the retarded and nonretarded regimes for generic systems \cite{Broer21,Lu2016,Munday2006}, however, due to the lack of decomposition of $s$ and $p$ modes, transparent solutions with clear asymptotic signatures are not available at present. The recent experimental demonstration of a Casimir torque between birefringent and liquid systems \cite{Somers2018} has given an impetus to further investigate this phenomenon, including in the context of new potential materials and more transparent expressions to better understand the limiting factor for the torque. The anisotropic TBG due to nematicity becomes a novel material platform to bring forward our understanding of Casimir torque.

\onecolumngrid
\begin{center}
\begin{figure}[ht]
    \includegraphics[width = 0.95 \textwidth]{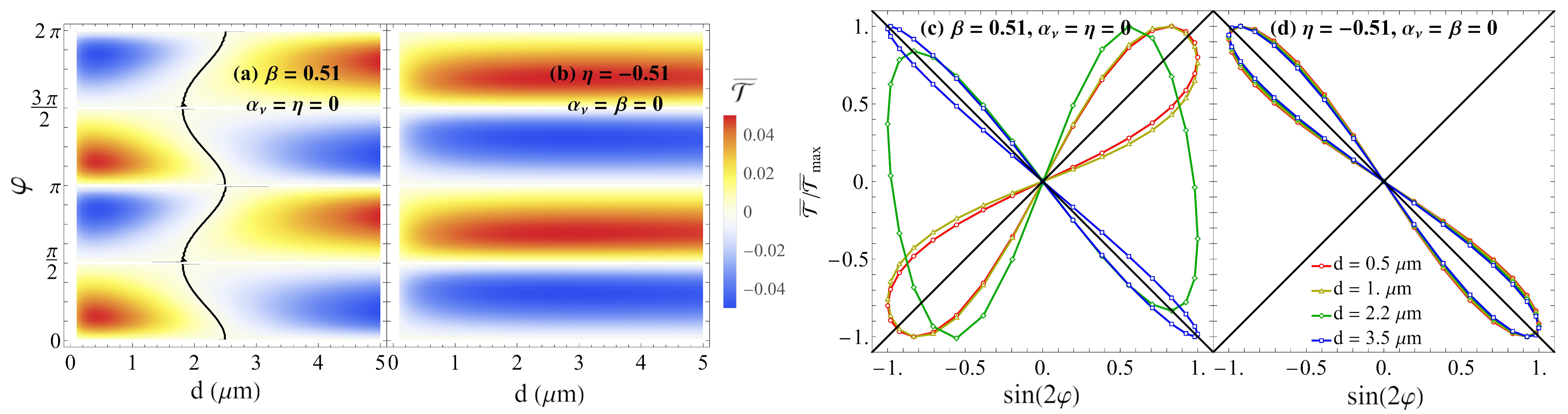}
    \caption{\label{fig:5} Density maps of $\overline{\mathcal{T}} = \frac{1}{E_{M}}\frac{\partial E(\varphi,d)}{\partial\varphi}$ in the relative angle $\varphi$ vs distance $d$ for: (a) TBGs with $\beta=0.51, \alpha_\nu=\eta=0$ and (b) TBGs with $\eta=-0.51, \alpha_\nu=\beta=0$. The ratio $\overline{\mathcal {T}}/\overline{\mathcal {T}}_{max}$, where $\overline{\mathcal {T}}_{max}$ denotes the maximum torque found for each distance $d$, as a function of $\sin (2\varphi)$ for (c) TBGs with $\beta=0.51, \alpha_\nu=\eta=0$ and (d) TBGs with $\eta=-0.51, \alpha_\nu=\beta=0$ for several separations. The black solid curve in (a) traces out the zeros of the function $\mathcal{H} (d, \varphi)$ on Eq. \eqref{eqn:2}.}  
\end{figure}
\end{center}
\twocolumngrid

\begin{center}
\begin{figure}[ht]
    \includegraphics[width = 0.95 \columnwidth]{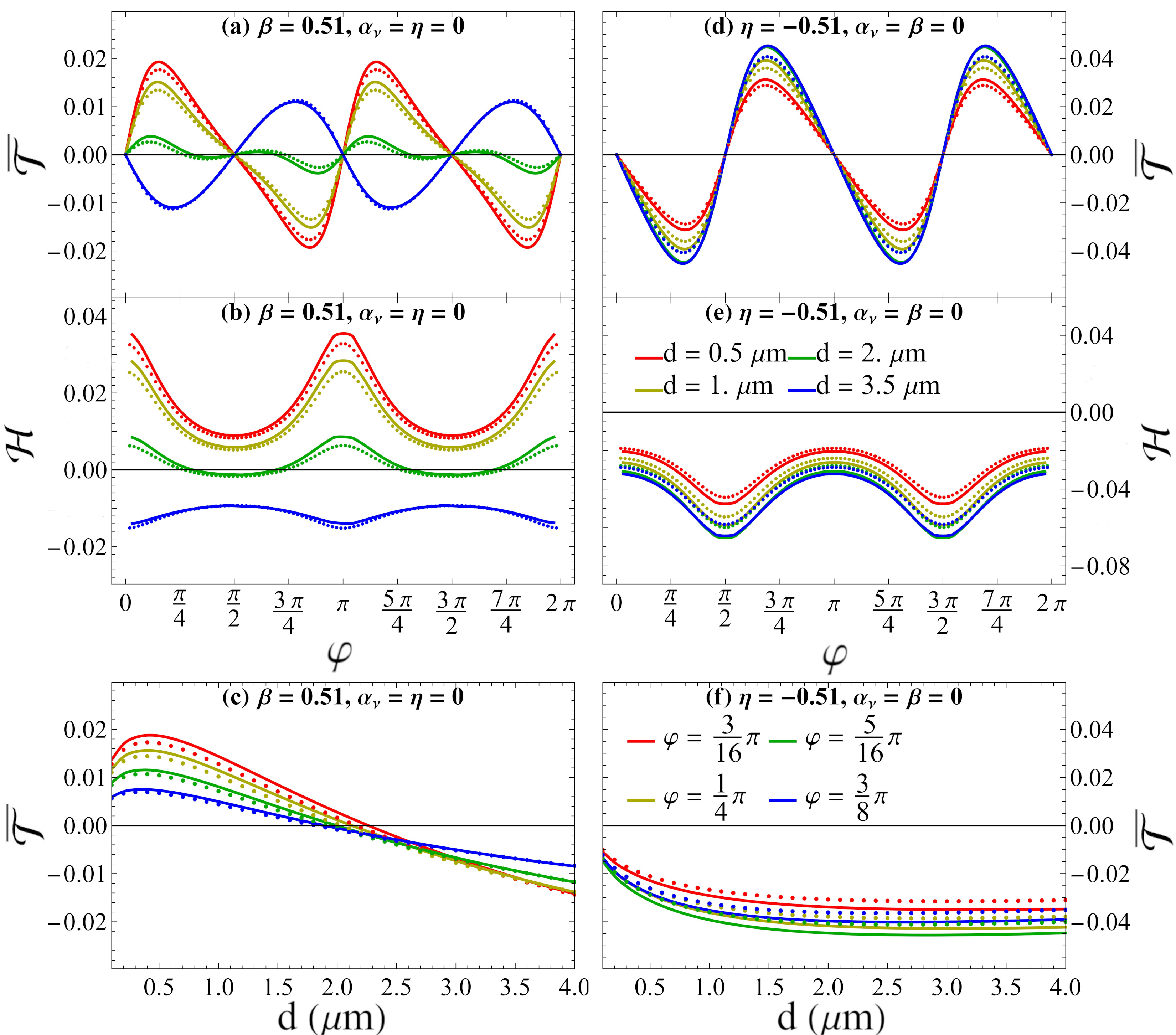}
    \caption{The normalized to $E_M$ Casimir torque $\overline{\mathcal{T}} = \frac{1}{E_{M}}\frac{\partial E(\varphi,d)}{\partial\varphi}$ as a function of the relative angle $\varphi$ for the TBG nematic state: (a) $\beta=0.51$ and (d) $\eta=-0.51$. The function $\mathcal{H}(d, \varphi)$ vs. the relative angle $\varphi$ for the TBG nematic state: (b) $\beta=0.51$ and (e) $\eta=-0.51$. The TBG distance separations are denoted in panel (e) for both $\overline{\mathcal{T}}$ and $\mathcal{H}$. The normalized to $E_M$ Casimir torque $\overline{\mathcal{T}} = \frac{1}{E_{M}}\frac{\partial E(\varphi,d)}{\partial\varphi}$ as a function of distance $d$ for several angles for TBG nematic states with: (c) $\beta=0.51$ and (f) $\eta=-0.51$. The legend for $\varphi$ shown in panel (f) applies for both states. The solid lines are obtained numerically from Eq. \eqref{eqn:7}, while the doted lines are obtained using Eq. \eqref{eqn:2}.}
    \label{fig:6}
\end{figure}
\end{center}

The Casimir torque is investigated next by calculating $\mathcal {T}=\frac{\partial (E(\varphi,d))}{\partial \varphi}$ for the nematic states of the TBGs.  In Fig. \ref{fig:5}, density plots of the re-scaled 
$\overline{\mathcal {T}}=\frac{\mathcal {T}}{E_M}$ are shown in a $(\varphi, d)$ coordinate map. The alternating minima and maxima, denoted in dark blue and red colors, are a consequence of the oscillatory $E(\varphi)$ dependence (also shown in Fig. \ref{fig:4}(c,d)). It is interesting that in the case of $\beta=0.51$ the pattern changes as the phase of the oscillations experiences a $180^{\circ}$ flip in the region $d\approx (2,3)$ $\mu$m. For the $\eta=-0.51$ nematic order, however, the alternating minima and maxima are preserved in the entire distance range shown. The covariance map $\left( \sin (2\varphi), \frac{\overline{\mathcal{T}}}{\overline{\mathcal{T}}_{max}} \right)$ in Fig. \ref{fig:5}c,d imprints the torque changes upon $\sin(2\varphi)$ functionality. For $\beta=0.51$ TBGs, the ratio $\frac{\overline{\mathcal{T}}}{\overline{\mathcal{T}}_{max}}$ follows $\sin(2\varphi)$ at separations $d<2$ $\mu$m, but at $d>3$ $\mu$m the torque ratio is consistent with $(-\sin(2\varphi))$. At intermediate distances, the {\it Lissajous}-like curve features the occurring phase transition (Fig. \ref{fig:5}c). For $\eta=-0.51$ TBGs, the torque follows strictly $-\sin(2\varphi)$, as evident from Fig. \ref{fig:5}d. 

To further understand the Casimir torque, we find that it can be represented using an approximate semi-analytical expression,
\begin{equation}
\label{eqn:2}
\mathcal{T} \approx -\frac{\pi^2 \hbar c}{720d^3}\mathcal{H} \left(d , \varphi \right) \sin(2 \varphi).
\end{equation}
It is obtained by neglecting the $\mathbb{R}^{(1)}\mathbb{R}^{(2)}$ term in the denominator of  Eq. \eqref{eqn:7} for $\mathcal{T}$ given in Methods of Calculations. Such an approximation is justified due to the small optical conductivity since all of its components essentially scale with $\alpha_{\rm fs}$. The function $\mathcal{H} \left(d , \varphi \right)$ is given in the Supplemental Information and it is expressed in terms of $\Delta \sigma=\sigma_{xx}-\sigma_{yy}$ and $\Sigma=\sigma_{xx}+\sigma_{yy}$. To verify the validity of Eq.\eqref{eqn:2}, in Fig. \ref{fig:6} we show the torque as a function of $\varphi$ for several separations and as a function of $d$ for different angles calculated completely numerically and using the semi-analytical expression given above. The excellent agreement gives confidence that Eq. \eqref{eqn:2} captures the Casimir torque in TBGs. 

From Eq. \eqref{eqn:2}, we note that in general $\mathcal{T} \sim \sin(2\varphi)$, which is consistent with similar oscillatory dependence in other systems \cite{Munday2006, Somers2018}. We also note that $\mathcal{H}\sim \Delta \sigma$, but since the interband conductivities along the $x$ and $y$ axes do not differ much, it is concluded that the torque is primarily determined by the difference in the Drude terms, $\Delta D$. Thus it is completely transparent in the expression that the Casimir torque is directly proportional to the TBG anisotropy. 

Nevertheless, the interplay between $\Delta D$ and $\sin (2\varphi)$ with the $d, \varphi$ dependence in $\mathcal{H}$ can give rise to a more complex torque behavior, as is the case for the $\beta=0.51$ phase change in the $d=(2,3)$ $\mu$m range. The graphical representation of $\mathcal{H}$ in Fig. \ref{fig:6}b shows an oscillatory behavior, whose maxima and minima are consistent with $\sin(2\varphi)$. For larger $d$ the amplitude is reduced and eventually the oscillation part of $\mathcal{H}$ becomes $\left(- \sin(2\varphi)\right)$ (as shown for $d=3$ $\mu$m). Thus, the overlap between these two types of oscillations is responsible for the phase change  in $\mathcal{T}$, which can also be seen in Fig. \ref{fig:6}c showing the torque changing sign as a function of $d$. In contrast, for the $\eta=-0.51$ nematic case, both $\mathcal{T}$ and $\mathcal{H}$ display similar oscillations preserving $\sin(2\varphi)$ behavior with a monotonic dependence of the torque as a function of separation, as evident in Fig. \ref{fig:6}d,e,f.

\section{Discussion}

TBGs at magic angles have a complex phase diagram near the CNP, which in turn gives diverse features in its optical response. Starting from a fully relaxed tight binding model, as the one given in \cite{Carr2019}, one can introduce phenomenologically various types of symmetry breaking states driven by electronic correlations \cite{Calderon2020}. The band structure and associated optical conductivity experience a rich structure in the order parameters $\alpha_{\nu}, \beta, \eta$ vs doping space showing different nematic orders and topologically nontrivial states that are relevant experimentally \cite{Po2018,Po2019,Choi2019,Xie2019,Lu2019,Stepanov2020}. 

Here, we have taken representative values of $\alpha_{\nu}, \beta, \eta$ parameters at the CNP to showcase how the different electronic phases and their consequences in the optical response determine the Casimir interaction and Casimir torque. Although there maybe quantitative differences, the qualitative behavior of the energy $E$ for each phase and the appearance of the torque ${\mathcal {T}}$ for the nematic states is robust. Similar phases are expected for TBGs at different filling factors \cite{Sharpe2019, Serlin2020, Xie2021}. Specifically, the broken $C_3$ symmetry effectively included via $\eta$ and $\beta$ parameters guarantees the anisotropy in the Drude weights, but the particular values of these parameters can quantitatively affect $D_{xx}-D_{yy}$, meaning that the magnitude of the torque may be changed.  Similarly, the nonzero gap at the Fermi level due to broken $T$ symmetry ensures the QAHE emergence leading to repulsive quantized Casimir interaction, while broken $C_2$ symmetry changes the $E$ vs $d$ dependence as compared to the noncorrelated TBG state. Nevertheless, given that this quantum mechanical interaction is valid at $k_BT/E_g<1$, the size of the gap determined by the particular value of $\alpha$ dictates the range of validity of this type of functionality.

The relaxation also affects the electronic structure and the value of the magic angle \cite{Fang2016,Gargiulo2017,Wijk2014, Koshino2018, Guinea2019}. In Fig. S-1b in the  Supplementary Information, we show the calculated band structure for a $\theta_{PR}=1.05^{\circ}$ TBG described with a partially relaxed (PR) model with relaxation included only along the vertical axis perpendicular to the bilayer \cite{Po2019}. We use representative examples for the correlation parameters 
$\alpha_\nu=2$ breaking the $C_2 T$ symmetry, and $\beta = -0.6$  and $\eta = -0.1$ breaking the $C_3$ symmetry. The corresponding band structures are displayed in Fig. S-2. Using the Kubo formalism, the optical response for the magic angle TBG within the partially relaxed model is also given in Fig. S-3 showing the same characteristics in imaginary frequency as described earlier for the fully relaxed model. As a result, the Casimir interaction and torque experience the same qualitative behavior as well, as shown in Fig. S-5, S-6, and S-7 in the Supplementary Information. Thus, the results for the Casimir phenomena are robust to the degree of relaxation in the TBG band structure.

The experimental observation of these unique Casimir interaction effects is possible in principle as they can serve as fingerprints of the different correlated TBG phases. Measuring the scaling behavior and comparing to the results in Fig. \ref{fig:4} can distinguish between the $C_2T$ and $C_3$ symmetry breaking phases. The presence of repulsion is an indication of a $T$-symmetry breaking state, while measuring non-zero torque, driven by Drude anisotropy, suggests nematicity.  The distinct phases of magic angle TBGs are low temperature phenomena, i.e below $4 \; K$ \cite{Saito2021, Wu2021}. Casimir forces as a function of distance and optical axis orientation for anisotropic materials have been measured at liquid He temperatures \cite{Norte2018,Wang2021,Laurent2012} indicating that such experiments of the phases of the magic angle TBGs are possible in the laboratory. Additionally, our calculations show that the Casimir torque for nematic TBGs is in the $6$ nN.m/m$^2$ to $0.01$ nN.m/m$^2$ range for the separation of $0.1 \mu$m to $1 \mu$m, which is achievable experimentally \cite{Somers2018}. Graphene Casimir forces are in the range of the ones shown here and they have been already measured as well \cite{Liu2021B,Mohideen2021}. 

Magic angle TBGs are a fertile ground for new states, which are largely driven by electronic correlations, including the various topological phases already demonstrated experimentally. Although much of this emergent research is focused primarily on the electronic structure properties, we show that Casimir interactions can also be used as effective means for probing, including the nontrivial topology and Drude anisotropy that can be supported by TBGs. These results further broaden our understanding of light-matter interactions in materials derived from graphene.

\section{Methods of calculations}

{\it Electronic Structure.} The fully relaxed model employed here follows the scheme given in \cite{Po2019,Carr2019}, which includes for each valley ten effective orbitals associated with the moir{\'e} patterns and their underlying symmetries. Specifically, we have $p_{zT}, p_{+T}, p_{-T}$ orbitals centered at the triangular lattice for the AA regions, $p_{+H}^A, p_{-H}^A, p_{+H}^B, p_{-H}^B$ at the hexagonal lattice formed by the AB and BA regions and $s_{\kappa1}, s_{\kappa2}, s_{\kappa3}$ at the kagome lattice of the domain walls. The energies of the two valleys are related by $\epsilon_{\overline{\nu}}(\bf{k})=\epsilon_{\nu}(-\bf{k})$. The $p_{+T}, p_{-T}$ are primarily responsible for the spectral weight of flat bands around the CNP and are believed to be strongly involved in the correlated states. The possible symmetry breakings are introduced phenomenologically via the parameters $\alpha_\nu$, $\beta$ and $\eta$ \cite{Choi2019,Calderon2020}.  $\alpha_\nu$ lifts the degeneracy between $p_{+T}$ and  $p_{-T}$, which in effect breaks the $C_2T$ symmetry by making the two 
sublattices in each layer inequivalent. Both $\beta, \eta$ break the rotational $C_3$ symmetry and are responsible for nematicity in the TBG: $\eta$ makes the interorbital $p_{+T}, p_{-T}$ hoppings inequivalent, while $\beta$ makes the interorbital hopping between $p_{+T}, p_{-T}$ and  $s_{\kappa 1}, s_{\kappa 2}, s_{\kappa 3}$ inequivalent. More details for the model are given in the Supplementary Information.

{\it Optical conductivity.} 
The optical conductivity tensor components are calculated within linear response by taking into account the coupling between electrons and electromagnetic fields via a Peierls substitution \cite{Valenzuela2013,Calderon2020}. The resulting expression for the conductivity tensor components is found as 

\begin{widetext}
\begin{eqnarray}
\text{Re}\left[\sigma_{\lambda\gamma}(\omega)\right]  &=& D_{\lambda\gamma}\delta(\omega) + \frac{1}{V}\sum_{{\bf{k}},l\neq l'} \Bigg\{ \Theta( \epsilon_{l',{\bf{k}}} ) \Theta(-\epsilon_{l,{\bf{k}}})\Bigg[\frac{\pi \text{Re}\left[\Omega_{\lambda \gamma}^{ll'} \right] }{\epsilon_{l',\bf{k}} - \epsilon_{l,\bf{k}}} \delta(\omega-\epsilon_{l',{\bf{k}}}+\epsilon_{l,{\bf{k}}}) \nonumber\\
&+&\left( \mathcal{P} \left( \frac{1}{ \epsilon_{l',\bf{k}} - \epsilon_{l,\bf{k}}-\omega} \right) - \mathcal{P}\left( \frac{1}{\epsilon_{l',\bf{k}} - \epsilon_{l,\bf{k}}+\omega} \right) \right) \Bigg] \Bigg\}
\label{eq:conductivity}
\end{eqnarray}
\end{widetext}
with $\mathcal{P}$ denotes the principal part. The $\text{Im}\left[\Omega_{\lambda \gamma}^{ll'}\right]$ term vanishes for the diagonal conductivity and it is responsible for the Hall conductivity at zero frequency if the total Chern number is finite. The Drude weight $D_{\lambda\gamma}$ includes diamagnetic and paramagnetic contributions is also obtained,

\begin{eqnarray}
D_{\lambda\gamma}=&& -\pi \sum_{kl} t^{\lambda\gamma}_{ll}\Theta(-\epsilon_{l,{\bf{k}}}) \nonumber\\
&& - \frac{2\pi}{V}\sum_{{\bf{k}},l\neq l'}\frac{\text{Re}\left[\Omega_{\lambda \gamma}^{ll'}\right]}{\epsilon_{l',\bf{k}} - \epsilon_{l,\bf{k}}}
\Theta( \epsilon_{l',{\bf{k}}} ) \Theta(-\epsilon_{l,{\bf{k}}}), \\
\Omega_{\lambda \gamma}^{ll'}=&&\sum_{\mu\nu\mu'\nu'} \frac{\partial \epsilon_{\mu\nu,\bf{k}}} {\partial k_\lambda} \frac{\partial \epsilon_{\mu'\nu',\bf{k}}}{\partial k_\gamma} a^*_{\mu,l,\bf{k}}a_{\nu,l',\bf{k}} a^*_{\mu',l',\bf{k}}a_{\nu',l,\bf{k}}, \\
t^{\lambda\gamma}_{ll}=&&\sum_{\mu\nu} \frac{\partial^2 \epsilon_{\mu \nu, \bf{k}}}{\partial k_\lambda \partial k_\gamma}a^*_{\mu l,\bf{k}} a_{\nu l,\bf{k}}.
\end{eqnarray}
Here $\epsilon_{\mu\nu,\bf{k}}$ denotes the bilinear terms of the Hamiltonian in the orbital basis, including the two valleys, $\epsilon_{l,\bf{k}}$ are the band energies, and $a_{\nu l,\bf{k}}$ changes from the band to the orbital basis \cite{Valenzuela2013,Calderon2020}. 

To obtain the optical response in imaginary frequency, we use the Kramers-Kronig relations valid for all real frequency to find the real part of the conductivity tensor components at imaginary frequencies $i\xi$,
\begin{eqnarray}
\label{eqn:5}
\text{Re} [\sigma_{\lambda\gamma}(i\xi)] && =\frac{2}{\pi}\int_0^\infty d\omega \frac{\xi}{\omega^2 +\xi^2} \text{Re}[\sigma_{\lambda\gamma}(\omega)] .
\end{eqnarray}
Since $\text{Im} [\sigma_{\lambda\gamma}(i\xi)]=0$, the above relation yields the full result for the conductivity components in imaginary frequency. By taking $\xi\rightarrow \xi+\gamma$ finite dissipation is also included, and here $\hbar \gamma = 0.01$ eV is used in all calculations.

{\it Casimir Energy and Torque.} The Casimir energy between two anisotropic planar objects separated by a distance $d$ along the $z$-axis (as is the case of TBGs in Fig. \ref{fig:1}) can be calculated using the Lifshitz formalism in the quantum mechanical limit by taking the optical response properties in the imaginary frequency domain $(i\xi)$. When the relative orientation of their optical axis is $\varphi$, the quantum mechanical limit of the energy per unit area can be written as  
\begin{eqnarray}
\label{eqn:6}
E = &&\hbar c \int_0^{\infty} \frac{d\kappa}{2\pi} \iint\limits_{\infty} \frac{d^2 \bf q}{(2\pi)^2} \ln \left\{ \det \left[1- \mathbb{R}^{(1)} ({\bf q}, i \kappa)\right. \right. \nonumber\\
&& \left. \left.  \times  e^{-d\sqrt{\kappa^2+{\bf q' }^{2}}} \mathbb{R}^{(2)}({\bf q}', i \kappa) e^{-d\sqrt{\kappa^2+{\bf q }^{2}}} \right] \right\} ,
\end{eqnarray}
where $\kappa =\xi/c$ and ${\bf q}$ is the 2D wave vector in the $xy$-plane. The 2D layer positioned at $z=0$ has its optical axis aligned along $x$, while the 2D layer at  $z=d$ is rotated by $\varphi$ around $z$. Therefore, ${\bf q'}=R_\varphi {\bf q}$ corresponds to the rotated wave vector with the rotation transformation matrix $R_\varphi=
\begin{pmatrix}
	\cos\varphi & -\sin \varphi \\
	\sin\varphi   &  \cos\varphi
\end{pmatrix}$. Consequently, the Casimir torque per unit area is given by
\begin{eqnarray}
\label{eqn:7}
&& \mathcal{T} (\varphi) =  - \hbar c \int_0^{\infty} \frac{d\kappa}{2\pi} \int \frac{d^2 \bf q}{(2\pi)^2} \nonumber\\
&& \quad \quad \times \text{tr} \left( \frac{\mathbb{R}^{(1)} (\mathbf{q}, i \kappa) \frac{d \mathbb{R}^{(2)} (\mathbf{q'}, i \kappa) }{d \varphi}}{ e^{2 d \sqrt{\kappa ^2 + \mathbf{q}^2}} - \mathbb{R}^{(1)} (\mathbf{q}, i \kappa) \mathbb{R}^{(2)} (\mathbf{q'}, i \kappa)  } \right).
\end{eqnarray}

The Fresnel reflection matrices $\mathbb{R}^{(j)}, j=1,2$ are obtained using Maxwell's equations and standard electromagnetic boundary conditions, 
\begin{eqnarray}
\mathbb{R}_{11}^{(j)} &=&-\frac{2\pi}{\Delta^{(j)}}\left[\frac{\sigma_{yy}^{(j)}}{\lambda c}+\frac{\det(\sigma ^{(j)})}{c^2} \right], \label{eqn:8}\\
\mathbb{R}_{12}^{(j)} &=&-\frac{2\pi}{\Delta^{(j)}} \frac{\sigma^{(j)}_{yx}}{c}, \label{eqn:9}\\
\mathbb{R}_{21}^{(j)} &=&\frac{2\pi}{\Delta^{(j)}} \frac{\sigma^{(j)}_{xy}}{c}, \label{eqn:10}\\
\mathbb{R}_{22}^{(j)} &=&\frac{2\pi}{\Delta^{(j)}}\left[\frac{\lambda \sigma_{xx}^{(j)}}{c}+\frac{\det(\sigma ^{(j)})}{c^2} \right], \label{eqn:11}\\
\Delta^{(j)} &=& 1 + 2 \pi \left[\frac{\lambda \sigma_{xx}^{(j)}}{c} + \frac{\sigma_{yy}^{(j)}}{\lambda c} \right] + \frac{4\pi^2}{c^2} \det(\sigma^{(j)}), \label{eqn:12}
\end{eqnarray}
where $\det(\sigma^{(j)})=\sigma^{(j)}_{xx}\sigma^{(j)}_{yy}-\sigma^{(j)}_{xy}\sigma^{(j)}_{yx}$ and $\lambda= \sqrt{{\bf q}^2 +\kappa^2}/\kappa$. Note further  that $\sigma^{(2)}=R_\varphi^{-1} \sigma^{(1)} R_\varphi$, which reflects the rotation of the 2D layer at $z=d$ by an angle $\varphi$. As a result, there are now off-diagonal contributions in the optical response which include not only the Hall conductivity $\sigma_{xy}$, but also the difference between the diagonal terms $\sigma_{xx,yy}$, which is essentially controlled by the Drude term anisotropy for the nematic orders.

\section*{Data availability}
All data supporting the findings of this study are available from the corresponding author upon reasonable request.

\section*{Acknowledgments}
L.M.W. acknowledges financial support from the US Department of Energy under grant No. DE-FG02-06ER46297. P. R.-L. was supported by AYUDA PUENTE 2021, URJC. M.J.C and E.B. acknowledge funding from PGC2018-097018-B-I00 (MCIN/AEI/FEDER, EU).

\section*{Author contributions}
P. R.-L. and D.-N. L. performed calculations for the Casimir interactions and torques. M. J. C and E. B. developed models for the electronic structure and optical response properties. L. M. W. conceived the idea, performed the analysis, and wrote the paper. 

\section*{Competing interests}
The authors declare no competing interest.

%

\pagebreak
\widetext
\begin{center}
\textbf{Supplementary Information: \\Twisted bilayered graphenes at magic angles and Casimir interactions: correlation-driven effects}\\
Pablo Rodriguez-Lopez $^{1}$, Dai-Nam Le $^{2,3}$, Mar\'{i}a J. Calder\'{o}n $^{4}$, Elena Bascones $^{4}$ and Lilia M. Woods $^{2, *}$ \\
$^1${\'A}rea de Electromagnetismo and Grupo Interdisciplinar de Sistemas Complejos (GISC), Universidad Rey Juan Carlos, 28933, M{\'o}stoles, Madrid, Spain\\
$^2$ Department of Physics, University of South Florida, Tampa, Florida 33620, USA\\
$^3$Atomic~Molecular~and~Optical~Physics~Research~Group, Advanced Institute of Materials Science, Ton Duc Thang University, Ho~Chi~Minh~City 700000, Vietnam\\
$^4$Instituto de Ciencia de Materiales de Madrid (ICMM), Consejo Superior de Investigaciones Científicas (CSIC), Sor Juana Inés de la Cruz 3, 28049 Madrid, Spain.\\
Emails: pablo.ropez@urjc.es (P. R.-L.), dainamle@usf.edu (D.-N. L.), calderon@icmm.csic.es (M. J. C.), leni.bascones@csic.es (E. B.), lmwoods@usf.edu (L. M. W., Corresponding author).
\end{center}
\setcounter{section}{0}
\setcounter{equation}{0}
\setcounter{figure}{0}
\setcounter{table}{0}
\setcounter{page}{1}
\makeatletter
\renewcommand{\thepage}{S-\arabic{page}} 
\renewcommand{\thesection}{S-\Roman{section}}  
\renewcommand{\thetable}{S-\Roman{table}}  
\renewcommand{\thefigure}{S-\arabic{figure}}
\renewcommand{\theequation}{S-\arabic{equation}}

\section{Electronic structure model of TBGs}

The non-interacting band structure is described with  a ten-band tight-binding model for each spin and valley based on effective moir{\'e} orbitals located at the triangular, hexagonal and kagome lattices formed by the symmetry points of TBG \cite{Po2019supp}: the triangular lattice for the AA regions is captured by $p_{zT}, p_{+T}, p_{-T}$ orbitals, the $p_{+H}^A, p_{-H}^A, p_{+H}^B, p_{-H}^B$ are associated with the AB and BA regions and their symmetry, while $s_{\kappa 1}, s_{\kappa 2}, s_{\kappa 3}$ at the kagome lattice of the domain walls. This tight-binding model has been used to fit an ab-initio  $k \cdot p$ continuum approach which considered a fully relaxed lattice with a twist angle $\theta_{FR}=0.9^\circ$ \cite{Carr2019supp}. A similar fitting model to a continuum theory for TBG with a twist angle of $\theta_{PR}=1.05 ^\circ$ was given in \cite{Po2019supp}, in which out of plane relaxation is taken into account. The terms of the ten-band tight-binding Hamiltonian and the parameters for both the fully and partially relaxed models can be found in the supplementary information of \cite{Calderon2020supp}. In the main text, we provide results based on the fully relaxed approach, while the partially relaxed method is consdered in the Supplementary Inormation for comparison. 

\subsection{Effective symmetry breaking parameters}
The symmetry breaking correlated states are introduced phenomenologically via three different terms: $H_\alpha$, which breaks the C$_2$T symmetry, and $H_\eta$ and $H_\beta$ which break the C$_3$ symmetry~\cite{Choi2019supp,Calderon2020supp} giving rise to a nematic state. Specifically, we have 

\begin{equation}
H_{\alpha}=\alpha_\nu \,\Sigma_{\bf k}( p^\dagger_{+,T,{\bf k}}p_{+,T,{\bf k}}-p^\dagger_{-,T,{\bf k}}p_{-,T,{\bf k}}) \, ,
\end{equation}

\begin{equation}
H_{\eta}=\eta \,\Sigma_{\bf k}(\phi_{01} +\phi_{10} +\phi_{-1-1}) p^\dagger_{+,T,{\bf k}}p_{-,T,{\bf k}}+h.c, 
\end{equation}

\begin{equation}
H_{\beta}=
\beta \,\Sigma_{\bf k}\left(s^\dagger_{\kappa_1,{\bf k}}  \,s^\dagger_{\kappa_2,{\bf k}}  \,  s^\dagger_{\kappa_3,{\bf k}}  \right) 
\begin{pmatrix}
-\phi_{0-\frac{1}{2}} &  -\phi_{0\frac{1}{2}}  \\
0.5\phi_{\frac{1}{2}\frac{1}{2}}e^{-i 2\pi/3} &  0.5\phi_{-\frac{1}{2}-\frac{1}{2}}e^{i 2\pi/3} \\
0.5 \phi_{-\frac{1}{2}0}e^{i 2\pi/3} & 0.5\phi_{\frac{1}{2}0}e^{-i 2\pi/3} 
\end{pmatrix}
\begin{pmatrix}
 p_{+,T,{\bf k}}  \\
 p_{-,T,{\bf k}} 
\end{pmatrix}
+ h.c..
\end{equation}
Here $\phi_{lm}= e^{-i {\bf k} \cdot (l {\bf a}_1+ m {\bf a}_2)}$ includes information about the hopping directions with ${\bf a}_1=a \hat y$ and ${\bf a}_2=a \left(\frac{\sqrt{3}}{2} \hat x -\frac{1}{2} \hat y \right)$ the lattice vectors, $l, m$ fractional numbers and $a$ the moir\'e lattice constant.

We distinguish the case of breaking the $C_2$ symmetry with $\alpha_\nu=\alpha_{\bar \nu}$ and the case of breaking the $T$ symmetry with $\alpha_\nu=-\alpha_{\bar \nu}$. The nematic parameters are not dependent on the valley degree of freedom.

\subsection{Band structure of TBGs in fully relaxed and partially relaxed models}

\begin{center}
\begin{figure}[H]
\centering
    \includegraphics[width = 0.6 \textwidth]{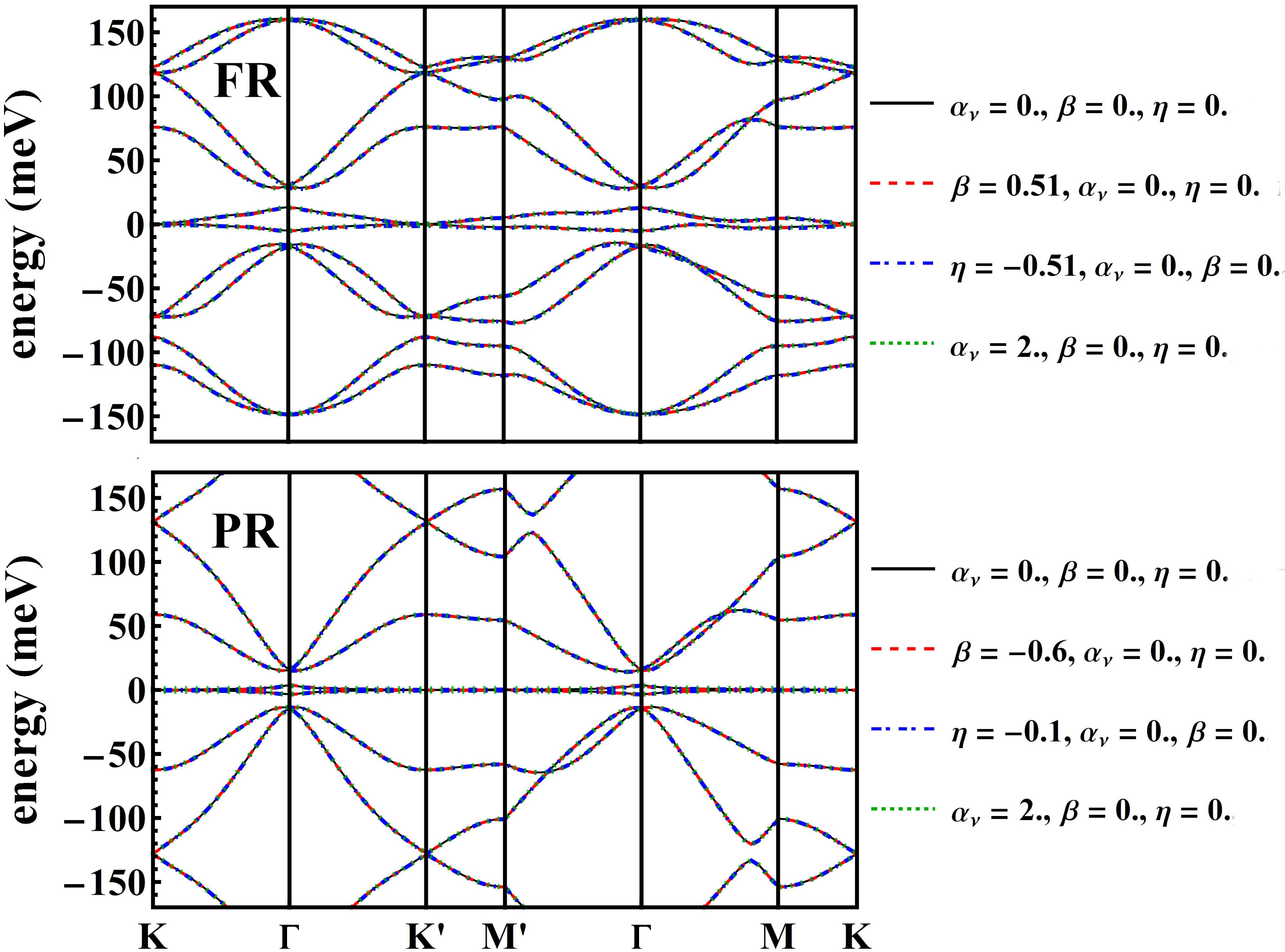}
    \caption{Full energy band structure for a single TBG for the two models considered. (Top panel) the ten-band fully relaxed model with a $\theta _{FR} = 0.9^{\circ}$ twist angle \cite{Carr2019supp} for the non-correlated state with $\alpha_\nu=\beta=\eta=0$ (black solid line),  for two nematic states breaking the $C_3$ symmetry: the  $\beta=0.51$ case (red dashed line), and the $\eta=-0.51$ one (blue dot-dashed line), and with $\alpha_{\nu} = \alpha_{\bar \nu} = 2$ breaking the $C_2T$ symmetry (green dotted line). (Bottom panel) the ten-band partially relaxed model with a $\theta _{PR} = 1.05^{\circ}$ twist angle \cite{Po2019supp} for non-correlated state (black solid line), $\beta=-0.6$ (red dashed line), $\eta=-0.1$ (blue dot-dashed line) and with $\alpha_{\nu} = \alpha_{\bar \nu} = 2$  (green dotted line).}
    \label{fig:S1}
\end{figure}
\end{center}

Fig. \ref{fig:S1} illustrates the band structure  obtained with the fully relaxed model for the $\theta _{FR} = 0.9^{\circ}$ TBG and with the partially relaxed model for the $\theta _{PR} = 1.05^{\circ}$ TBG in the non-correlated state and the three correlated states discussed in the main text. At the large scale of the ten bands considered, the differences between the bands are small. 

\begin{center}
\begin{figure}[H]
    \centering
    \includegraphics[width = 0.9 \textwidth]{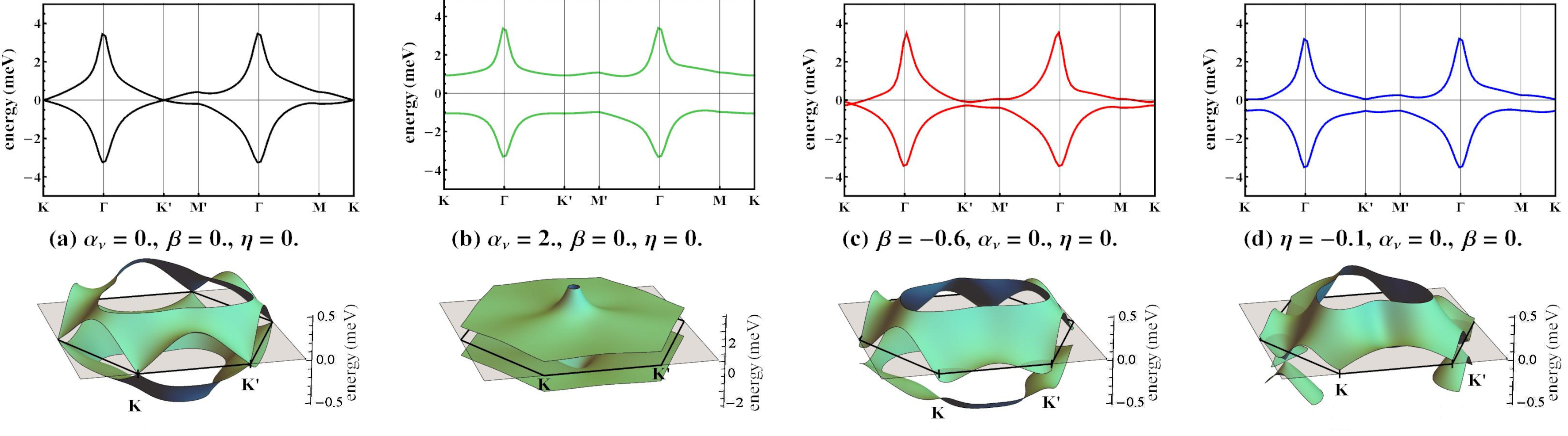}
    \caption{The low energy band structure for a single TBG with $\theta_{PR} = 1.05 ^\circ$ twist angle is given as calculated with the ten-band partially relaxed model \cite{Po2019supp}. The top panels show the bands along high symmetry directions and the bottom ones a zoom-in around the Fermi level in the whole Brillouin zone. (a) Non-correlated state with $\alpha_\nu=\beta=\eta=0$; (b) a state breaking the $C_2T$ symmetry; a nematic state with (c) $\beta=-0.6$ and (d) $\eta=-0.1$.} 
    \label{fig:S2}
\end{figure}
\end{center}

Fig. \ref{fig:S2} shows the flat bands around the charged neutrality point within the partially relaxed model \cite{Po2019supp} for the noncorrelated TBG state and three additional correlated states with parameters corresponding to the symmetry breaking ones in Fig. S-1. The 3D bandstructures are also shown under each panel.

\section{Conductivity in imaginary frequency for the partially relaxed model}

\begin{figure}[ht]
    \centering
    \includegraphics[width = 0.60 \textwidth]{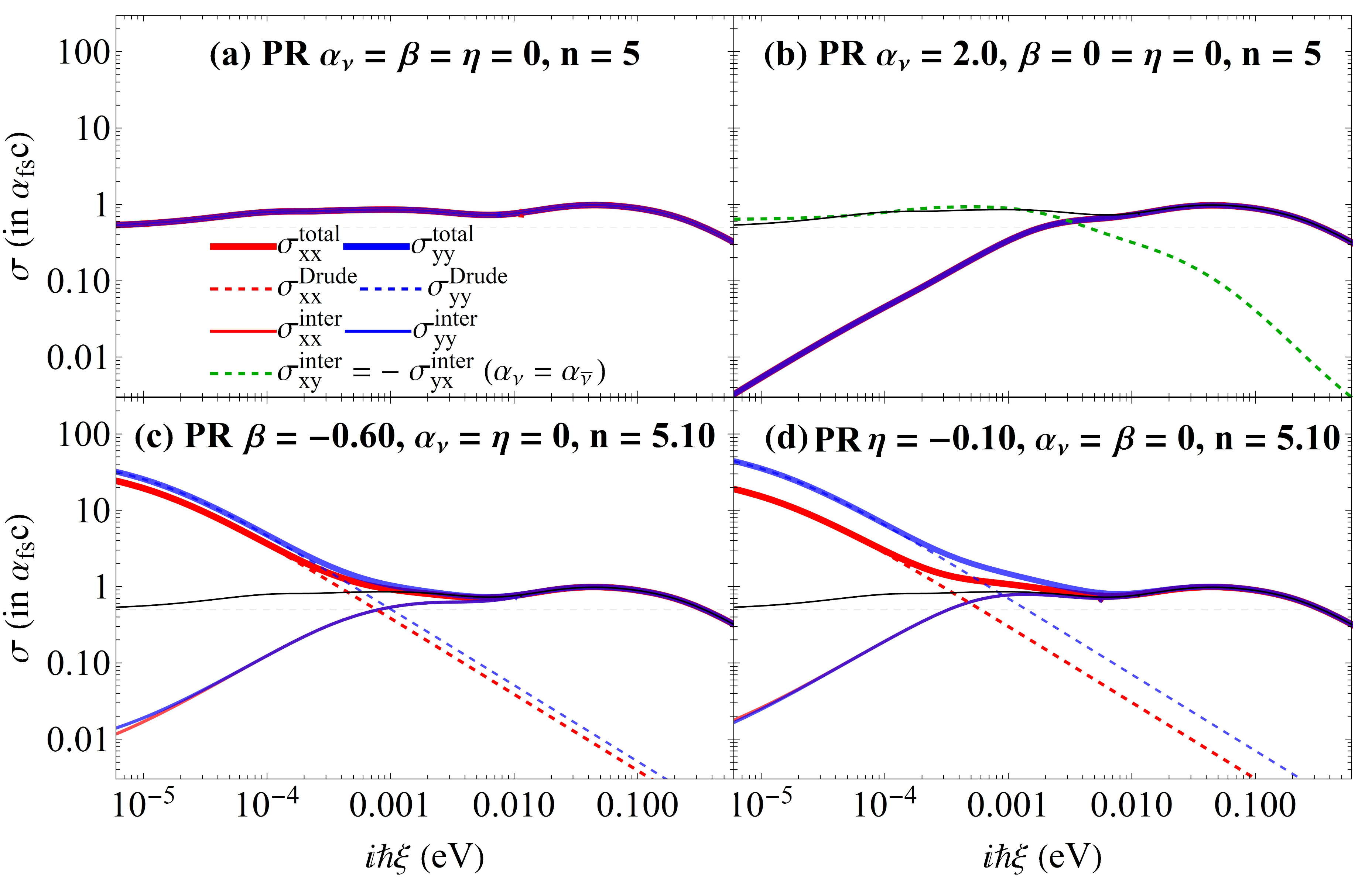}
    \caption{Optical conductivity components of a TBG with a twist angle $\theta_{PR} = 1.05^{\circ}$ for the partially relaxed model \cite{Po2019supp} for a: (a) non-correlated $\alpha_\nu=\beta=\eta=0$ state with all inherent symmetries intact; (b) insulating state with $\alpha_{\nu}=\alpha_{\bar \nu}=2$; a nematic state with (c) $\beta=-0.6$ and (d) $\eta=-0.1$. The thin black curve in (b, c, d) shows the diagonal optical conductivity for the non-correlated state as a reference.}
    \label{fig:S3}
\end{figure}

The optical conductivity tensor for the magic angle TBG with $\theta_{PR} = 1.05^{\circ}$ is calculated as detailed in the Methods section in the main text. In Fig. \ref{fig:S3} we show its components in the imaginary frequency domain. By comparing with Fig. 3 in the main text, it is clear that the behavior of $\sigma_{ij}\left(i\xi \right)$ is very similar for both models.

\section{Semi-analytical approximation for Casimir torque between anisotropic TBGs}

Utilizing the Jacobi formula $ \frac{d}{d \varphi} \ln \left[ \det\left( A (\varphi) \right) \right] = \tr\left( A^{-1} (\varphi) \frac{d A(\varphi) }{d \varphi} \right)$, one may obtain the Casimir torque from the interaction energy,
\begin{eqnarray}
\mathcal{T} (\varphi) = \frac{d E}{d \varphi} = - \hbar c \int_0^{\infty} \frac{d\kappa}{2\pi} \int \frac{d^2 \bf q}{(2\pi)^2} \tr \left( \frac{\mathbb{R}^{(1)} (\mathbf{q}, i \kappa) \frac{d \mathbb{R}^{(2)} (\mathbf{q'}, i \kappa) }{d \varphi}}{ e^{2 d \sqrt{\kappa ^2 + \mathbf{q}^2}} - \mathbb{R}^{(1)} (\mathbf{q}, i \kappa) \mathbb{R}^{(2)} (\mathbf{q'}, i \kappa)  } \right).
\end{eqnarray}
For large separation $d$ between two TBGs,
the exponential term in the denominator becomes dominant and the above expression can be approximated as
\begin{eqnarray}
\mathcal{T} (\varphi) \approx - \hbar c \int_0^{\infty} \frac{d\kappa}{2\pi} \int \frac{d^2 \bf q}{(2\pi)^2} \tr \left( e^{-2 d \sqrt{\kappa ^2 + \mathbf{q}^2}} \mathbb{R}^{(1)} (\mathbf{q}, i \kappa) \frac{d \mathbb{R}^{(2)} (\mathbf{q'}, i \kappa) }{d \varphi}\right).
\end{eqnarray}
After defining $\kappa = \frac{t u}{d} $ and $\mathbf{q} = \frac{t}{d} \sqrt{1-u^2} \hat{\mathbf{q}}$ where $t \in \left[0, + \infty \right)$ and $u = \frac{1}{\lambda} = \frac{\kappa}{\sqrt{\kappa^2+ \mathbf{q}^2}} \in \left[0, 1\right]$, the torque is given as 
\begin{eqnarray}
\mathcal{T} (\varphi) \approx - \frac{\pi^2 \hbar c }{720 d^3} \mathcal{H} \left(d , \varphi \right) \sin 2 \varphi,
\end{eqnarray}
where $\mathcal{H} (d, \varphi)$ is defined via the following integral
\begin{eqnarray}
\mathcal{H}(d, \varphi) = \frac{180}{\pi^{4}}
 \int_{0}^{\infty}t^{2}e^{-2t}dt \int_{0}^{1}\frac{ \Delta \sigma \left(A_{0} + A_{1}u + A_{2}u^{2} + A_{3}u^{3} + A_{4}u^{4} \right)}{ \left( 1 + B_{1}u \right) \left( u + B_{2} \right) \left( B_{3}u^{2} + B_{5} u + B_{4} \right)^{2}}udu,
\end{eqnarray}
with
\begin{eqnarray}
\left\{
\begin{array}{ccl}
A_{0} & = & \Delta\sigma + \Sigma\\
A_{1} & = & - ( \Delta\sigma + \Sigma )\left( \Delta\sigma^{3} - \Sigma \Delta\sigma^{2} - \Sigma^{2}\Delta + \Delta\sigma + \Sigma^{3} - 3\Sigma \right)\\
A_{2} & = & - 2\Delta\sigma \left( \Delta\sigma^{2} - \Sigma^{2} \right)\\
A_{3} & = & \left( \Delta\sigma - \Sigma \right) \left( \Delta\sigma ^3 + \Sigma \Delta\sigma ^2 - \Sigma^2 \Delta\sigma + \Delta\sigma - \Sigma^3 + 3 \Sigma \right) \\
A_{4} & = & \Delta\sigma - \Sigma \\
\end{array} \right., \quad \left\{
\begin{array}{ccl}
B_{1} & = & \Sigma - \Delta \sigma\\
B_{2} & = & \Sigma + \Delta \sigma\\
B_{3} & = & \Sigma - \Delta \sigma\cos(2\varphi)\\
B_{4} & = & \Sigma + \Delta \sigma\cos(2\varphi)\\
B_{5} & = & 1 + \Sigma^{2} - \Delta \sigma^{2}
\end{array} \right..
\end{eqnarray}
The above terms are related to the conductivity of TBG in imaginary frequency via
\begin{eqnarray}
&& \Sigma = \frac{\pi}{c}\left(\sigma_{xx} + \sigma_{yy}\right) = 2 \sigma^{inter} \left(i \frac{tu \hbar c}{d}\right) + \frac{\pi}{c} \left( \frac{D_{xx} + D_{yy}}{\frac{t u \hbar c}{d} + \hbar \Gamma } \right), \\
&& \Delta \sigma = \frac{\pi}{c}\left(\sigma_{xx} - \sigma_{yy}\right) = \frac{\pi}{c} \left( \frac{\Delta D}{\frac{t u \hbar c}{d} + \hbar \Gamma } \right).
\end{eqnarray}
Given that the interband optical anisotropy is much smaller than the one for the Drude terms, here we have taken that $\sigma_{xx}^{inter} \approx \sigma_{yy}^{inter} \approx \sigma^{inter}$.

Normalizing the Casimir torque by the Casimir interaction between two perfect metals $E_M = - \frac{\pi^2 \hbar c}{720 d^3}$ results in the reduced Casimir torque written as,
\begin{equation}
\overline{\mathcal{T}} (\varphi) = \frac{\mathcal{T}(\varphi)}{E_M} = \mathcal{H} \left(d , \varphi \right) \sin 2 \varphi.
\end{equation} 

\section{Casimir interactions between two TBGs with $\theta_{PR} = 1.05^{\circ}$}

Following the formalism described in the main text, we now calculate the Casimir energy and torque for the TBG with $\theta_{PR} = 1.05^{\circ}$. In the case of non-correlated state where $\alpha_\nu = \beta = \eta = 0$, in both fully relaxed and partially relaxed model of TBGs, the zero-frequency optical conductivity $\sigma_0$ becomes dominant in the long-range asymptotic behavior of the Casimir energy $E$. Since the zero-frequency optical conductivity of the TBG is twice that of the monolayer graphene $\sigma _0 \approx 2 \sigma _g$, the interaction energy per unit of area between two identical TBGs is twice the one between two identical monolayer graphene $E\approx 2E_g$ for large $d$. Fig. \ref{fig:S4} below shows how the ratio between Casimir energy $E$ of two non-correlated TBGs and Casimir energy $E_g$ of two monolayer graphene evolves as a function of separation distance $d$. This ratio approaches $\frac{E}{E_g}\approx 2$ as $d$ increases. The figure also illustrates that both results for the fully relaxed and partially relaxed models are very similar.

\begin{figure}[H]
    \centering
    \includegraphics[width = 0.39 \textwidth]{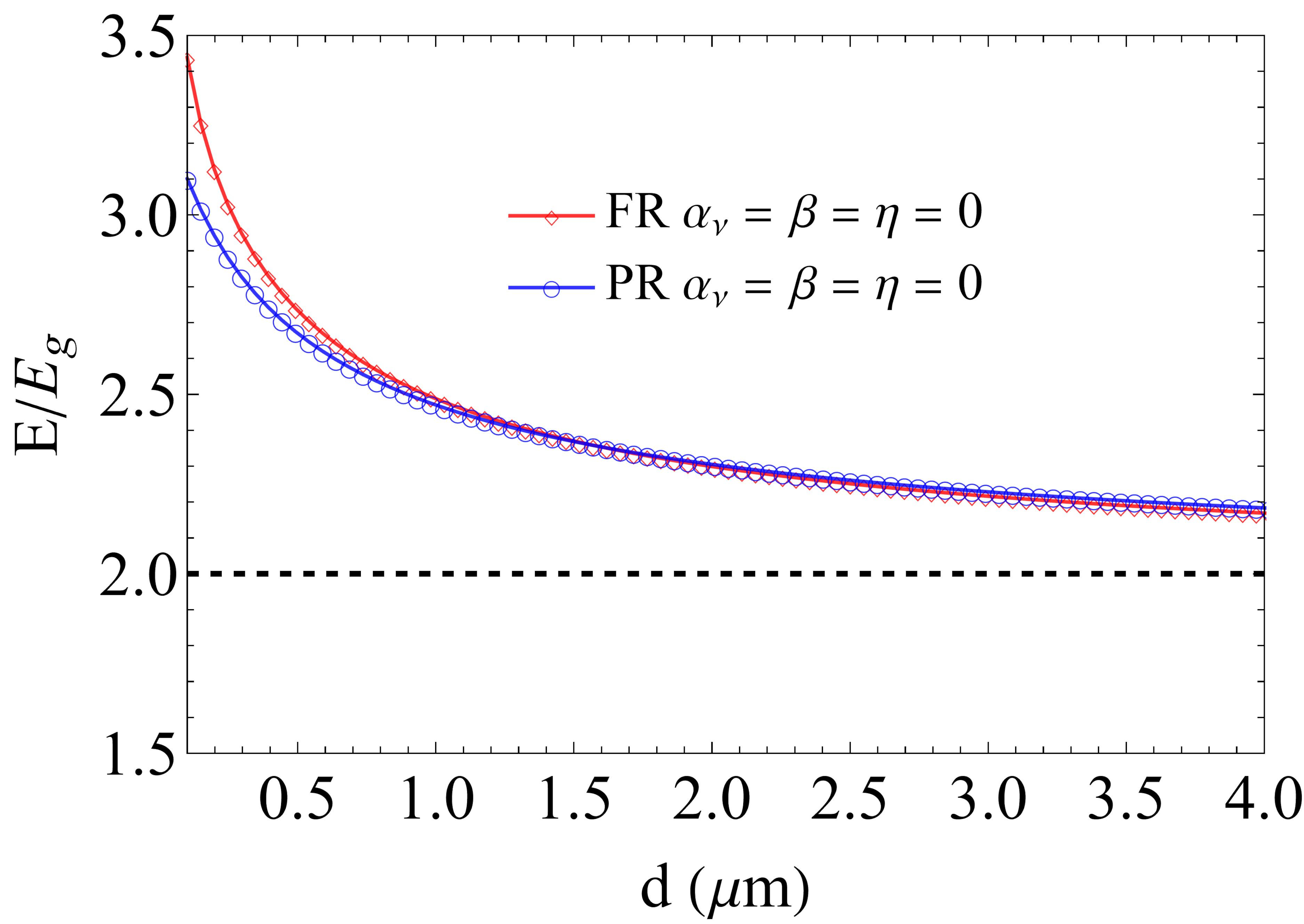}
    \caption{Casimir energy $E$ between two TBGs with non-correlated state $\alpha_\nu = \beta = \eta = 0$ for fully relaxed (FR) \cite{Carr2019supp}  $\theta_{FR} = 0.9^{\circ}$ (red line) and partially relaxed (PR) \cite{Po2019supp}  $\theta_{PR} = 1.05^{\circ}$ (blue line) models in comparison with double of Casimir energy between two graphene monolayers $2E_g$ (dashed line).}
    \label{fig:S4}
\end{figure}

\begin{figure}[H]
    \centering
   \includegraphics[width = 0.60 \textwidth]{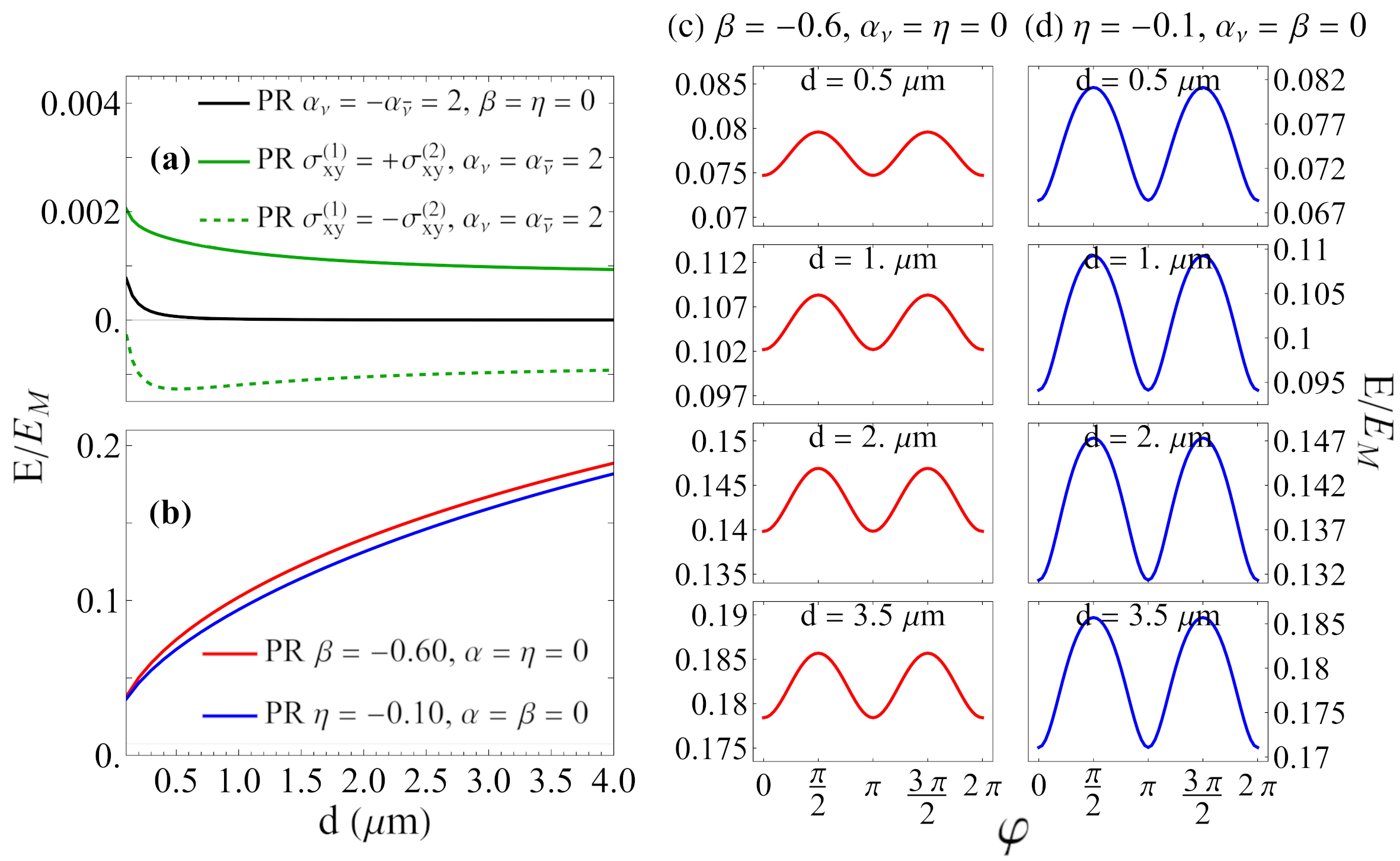}
    \caption{Casimir interaction energy $E$ between two identical TBGs normalized to the Casimir energy between two perfect metals $E_M=-\frac{\pi^2\hbar c}{720 d^3}$ for the partially relaxed model \cite{Po2019supp} $\theta_{PR} = 1.05^{\circ}$ for: (a) $\alpha_{\nu}=-\alpha_{\bar \nu}=2$ when the Hall conductivities of the two TBGs have the same (full green) and or opposite (dashed green) signs. The black curve corresponds to $\alpha_{\nu}=\alpha_{\bar \nu}=2$; (b) nematic states with $\beta=-0.6, \eta=-0.1$ parameters with aligned optical axes $\varphi=0$. The $E/E_M$ ratio as a function of the relative angle $\varphi$ between the TBG optical axes at several distances for: (c) $\beta=-0.6$ and (d) $\eta=-0.1$.}   
    \label{fig:S5}
\end{figure}

Fig. \ref{fig:S5} gives the numerical results for the Casimir energy for the different correlated states as a function of distance and angle of the optical axes of the $\theta_{PR} = 1.05^{\circ}$ TBG. One finds that the results exhibit the same qualitative and very close quantitative behavior as conpared to Fig. 4 in the main text.

\begin{center}
\begin{figure}[H]
\centering
    \includegraphics[width = 0.94 \textwidth]{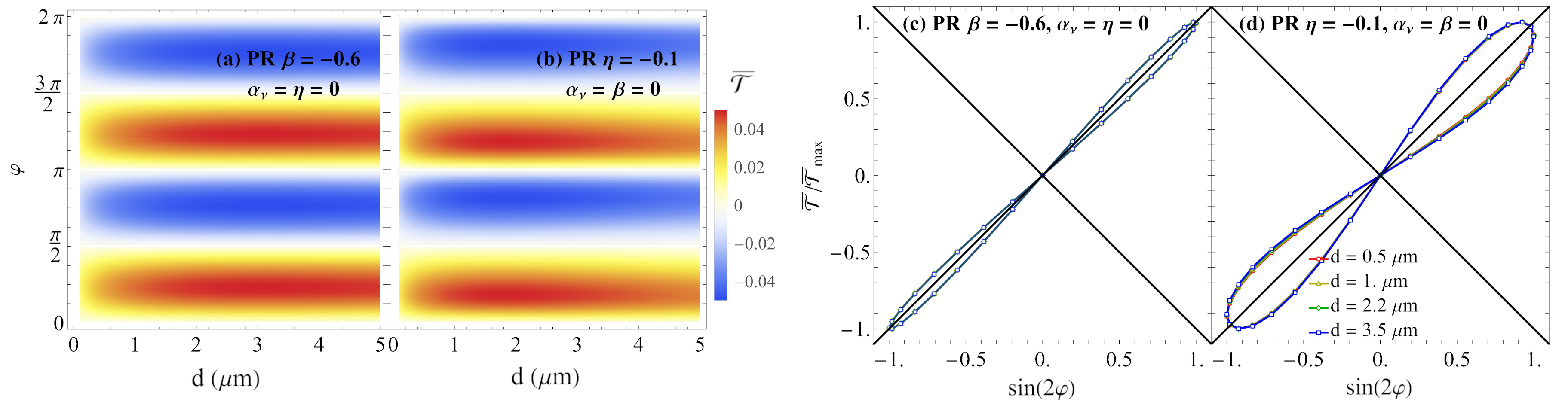}
    \caption{\label{fig:S6} Density maps of $\overline{\mathcal {T}}=\frac{\partial (E(\varphi,d)/E_M)}{\partial \varphi}$ in the relative angle $\varphi$ vs distance $d$ for: (a) TBGs in partially relaxed model $\theta_{PR} = 1.05^{\circ}$ with $\beta=-0.6, \alpha=\eta=0$ and (b) TBGs in partially relaxed model with $\eta=-0.1, \alpha=\beta=0$. The ratio $\overline{\mathcal {T}}/\overline{\mathcal {T}}_{max}$, where $\overline{\mathcal {T}}_{max}$ denotes the maximum torque found for each distance $d$, as a function of $\sin (2\varphi)$ for (c) TBGs in partially relaxed model with $\beta=-0.6, \alpha=\eta=0$ and (d) TBGs in partially relaxed model with $\eta=-0.1, \alpha=\beta=0$ for several separations.}  
\end{figure}
\end{center}

\begin{figure}[H]
    \centering
    \includegraphics[width = 0.65 \textwidth]{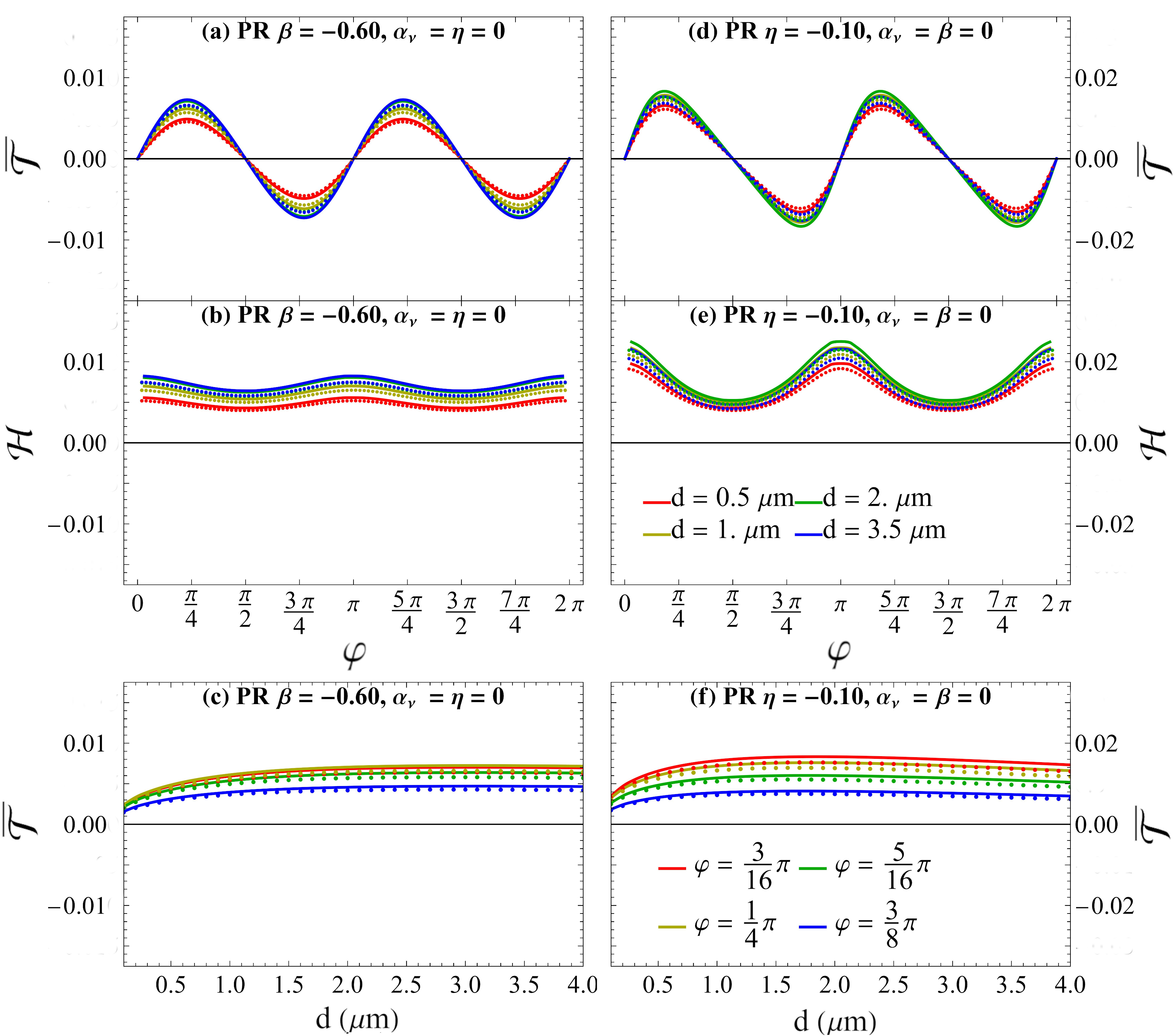}
    \caption{The normalized to $E_M$ Casimir torque $\overline{\mathcal {T}}=\frac{\partial (E(\varphi,d)/E_M)}{\partial \varphi}$ and the $\mathcal{H}$ function corresponding to a TBG with $\theta_{PR} = 1.05^{\circ}$ as functions of (a,b) relative angle $\varphi$ for several distances and (c) distance $d$ for several angles $\varphi$ for TBGs with $\beta=-0.6, \alpha_\nu=\eta=0$; (d, e) relative angle $\varphi$ for several separations and (f) distance $d$ for several angles $\varphi$ for TBGs with $\eta=-0.1, \alpha_\nu=\beta=0$. The solid lines are fully numerical obtained by Eq. (9) while the dots are semi-analytically obtained by (2).}
    \label{fig:S7}
\end{figure}

Results for the Casimir torque for the TBG with $\theta_{PR} = 1.05^{\circ}$ obtained within the partially relaxed model for the considered nematic states are shown in Fig. \ref{fig:S6} and Fig. \ref{fig:S7}. We find that for both nematic phases the torque is consistent 
with $\sin (2\varphi)$ periodic oscillations without phase change patterns, which appear in the $\beta=0.51$ case for the TBG with $\theta_{FR} = 0.9^{\circ}$ shown in Fig. 5 and 6 of the main text.
The semi-analytical expression highlighting the role of the Drude anisotropy is also confirmed for the TBG with $\theta_{PR} = 1.05^{\circ}$ within the partially relaxed model (Fig. \ref{fig:S7}).

\end{document}